\documentclass[screen, authorversion, acmsmall, nonacm]{acmart}

\usepackage{graphicx}
\usepackage{array}
\usepackage{float}
\usepackage{hyperref}
\usepackage{tabularx}
\usepackage{adjustbox}
\usepackage{booktabs}
\usepackage{natbib}
\usepackage{xcolor}
\usepackage{colortbl}
\usepackage{longtable}
\usepackage{multirow}
\usepackage{rotating}
\usepackage{enumitem}
\usepackage{amssymb}
\usepackage{amsmath}
\usepackage{pifont}
\usepackage{enumitem}
\usepackage{tablefootnote}
\usepackage{threeparttable}
\usepackage{appendix}
\usepackage{caption}
\usepackage{etoolbox}
\usepackage{fancyhdr}

\definecolor{researchdevelopment}{HTML}{1c78ba}
\definecolor{economy}{HTML}{0d6b70}
\definecolor{responsibleai}{HTML}{872996}
\definecolor{education}{HTML}{89D0E5}
\definecolor{diversity}{HTML}{eaadff}
\definecolor{policygovernce}{HTML}{4d1f52}
\definecolor{publicopinion}{HTML}{9cf2f2}
\definecolor{infrastructure}{HTML}{172e5c}

\makeatletter
\let\@authorsaddresses\@empty
\makeatother

\begin{document}

\thispagestyle{empty}

\vspace*{-1in}

\newcommand{\customtitle}[1]{
    \vspace{1cm} 
    \hrule height 1pt 
    \vspace{0.4cm} 
    \begin{center}
        \Large\textsc{#1}
    \end{center}
    \vspace{0.4cm} 
    \hrule height 1pt 
    \vspace{1cm} 
}
\customtitle{\textbf{The Global AI Vibrancy Tool} \\ November 2024}

\begin{center}
    \begin{tabular}{ccc}
        Loredana Fattorini & Nestor Maslej & Raymond Perrault \\
        \\
        Vanessa Parli & John Etchemendy & Yoav Shoham \\
       \\ & Katrina Ligett & \\
       \\
        & \textit{AI Index Project} & \\
        \\
        & \textit{Institute for Human-Centered AI, Stanford University} & \\
    \end{tabular}
\end{center}

\vspace{1cm}

\renewenvironment{abstract}{
    \if@twocolumn
      \section*{\abstractname}
    \else
      \begin{center}%
        {\bfseries\MakeUppercase{\abstractname}}
      \end{center}%
      \quotation
    \fi}
    {\if@twocolumn\else\endquotation\fi}

\begin{abstract}

This paper presents the latest version of the Global AI Vibrancy Tool (GVT), an interactive suite of visualizations designed to facilitate the comparison of AI vibrancy across 36 countries, using 42 indicators organized into 8 pillars. The tool offers customizable features that allow users to conduct in-depth country-level comparisons and longitudinal analyses of AI-related metrics, all based on publicly available data. By providing a transparent assessment of national progress in AI, it serves the diverse needs of policymakers, industry leaders, researchers, and the general public. Using weights for indicators and pillars developed by AI Index's panel of experts and combined into an index, the Global AI Vibrancy Ranking for 2023 places the United States first by a significant margin, followed by China and the United Kingdom. The ranking also highlights the rise of smaller nations such as Singapore when evaluated on both absolute and per capita bases. The tool offers three sub-indices for evaluating Global AI Vibrancy along different dimensions: the Innovation Index, the Economic Competitiveness Index, and the Policy, Governance, and Public Engagement Index.

\end{abstract}

\vspace{1cm}
\section{Introduction}

This paper introduces the latest version of the Global AI Vibrancy Tool (GVT), a comprehensive collection of publicly available time series data and a suite of interactive visualizations that allow comparisons for up to 36 countries across 42 indicators related to Artificial Intelligence (AI). This update expands on previous versions, covering more countries and metrics. It now features one of the most extensive collections of AI-specific indicators\footnote{In this framework, the terms ``indicator'' and ``metric'' are used interchangeably.} available, collectively assessing country-specific AI activity from 2017 to 2023. The revamped tool enhances the user experience with intuitive navigation, customizable visualizations, and download options. Additionally, the tool is structured to provide detailed and nuanced comparisons of countries through a user-customizable index of indicators in the Global and National AI Vibrancy Rankings section. The new GVT also offers an in-depth look at the evolution of specific indicators over time for selected countries in the AI Metrics over Time section. This new launch introduces additional AI vibrancy sub-indices, including the Innovation Index, the Economic Competitiveness Index, and the Policy, Governance, and Public Engagement Index.

In an era of rapid advancements in AI, various stakeholders require reliable metrics to gauge national progress in AI development. Policymakers need these metrics to build AI capacity by identifying effective policies, necessary research and development investments, and understanding the geopolitical impact of AI. Industry leaders seek insights for investment and strategy, aiming to understand which countries are emerging as AI hubs and where resources should be allocated. In addition, the general public may be interested in understanding how their country compares in AI activity and gaining insight into which nations are leading AI.

Existing AI national tracker tools often focus on narrow aspects such as investment levels or publication counts and may include broader, not necessarily AI-specific, indicators like the proportion of the population using the Internet or the number of STEM graduates. Although these metrics are valuable for understanding the general technological and educational infrastructure, they may not directly measure a country's AI-related progress and capabilities. The GVT addresses this gap by providing a holistic and interactive analysis platform that consolidates various AI progress indicators into a single, user-friendly interface, thereby promoting greater transparency, accountability, and knowledge sharing in the AI field. The design of the GVT involves a strategic selection of indicators organized into pillars and an overarching index, reducing AI metrics complexity. Users can also adjust indicator and pillar weights to reflect their own perspective.

The GVT will continually evolve, with future editions potentially incorporating new metrics and expanding country coverage. These changes will ensure that the tool remains a relevant and valuable resource for tracking global AI progress now and in the future.

The tool comes with a default set of weights for the indicators and pillars determined in consultation with a panel of experts. Using these weights, the United States has consistently held the top global position in AI vibrancy since 2018, maintaining a substantial lead over other nations and excelling across most dimensions, particularly in R\&D, Infrastructure, and Economy. China and the United Kingdom follow, with China showing strength in R\&D and Infrastructure, while the United Kingdom leads in Education as well as Policy and Governance. Interestingly, smaller nations like Singapore emerge as leaders in the ranking especially when adjusting the indicators on a per capita basis. These global rankings illustrate the diverse approaches countries are adopting to fostering AI growth, with both large and small nations making significant strides.

The remainder of this paper is organized as follows: In the \nameref{sec:related-works} section we review existing literature and tools related to AI metrics and comparisons. The \nameref{sec:conceptual-framework} section outlines the key concepts underlying the tool. The \nameref{sec:methodology} section details the processes involved in data collection, construction of the AI Vibrancy Index, and implementation within the tool. 
This is followed by a section providing a snapshot of the \nameref{sec:discussion}. 
Finally, the \nameref{sec:conclusion} provides a summary of the contributions and suggests directions for future improvements.

\section{Related Work}
\label{sec:related-works}

The development and refinement of composite indicators involve aggregating multiple individual metrics into a single comprehensive measure. This process draws on a diverse set of methodologies from statistics, economics, and technology assessment. Notable examples include the Human Development Index (HDI), introduced by the United Nations in 1990 \cite{united_nations_human_1990}, which integrates health, education, and income levels, and the Gender Inequality Index (GII), which measures gender disparities in health, empowerment, and labor market participation \cite{undp_united_nations_development_programme_hdr_2010}.
As the AI landscape continues to evolve rapidly, the need for robust tools to measure and compare the AI capabilities of different countries has become increasingly evident.

The \citet{oecd_handbook_2008} have laid down significant foundational work in the study of composite index creation. Their guidance stresses the importance of a coherent theoretical structure, meticulous data selection and transformation, and robustness checks to ensure the reliability of these indicators. Similarly, \citet{nardo_tools_2005} discuss the potential pitfalls of selecting, normalizing, weighting, and aggregating indicators. They emphasize the importance of transparency and consistency in the methodology, which are crucial for enhancing the interpretability and comparability of composite indices.

More recent advances in the methodological framework for composite indicators are discussed in the comprehensive review by \citet{greco_methodological_2019}. This review highlights the evolution in the adoption and methodological refinement of composite indicators due to their increased popularity in various research fields. \citet{greco_methodological_2019} specifically focused on important aspects, such as weighting and aggregation, areas that attract substantial criticism and suggest avenues for future research. Their work explores the robustness analysis that follows the construction of these indicators, a less explored but significant phase, highlighting the need for robust methodologies that can withstand scrutiny and provide reliable and interpretable results.

In addition to foundational knowledge, the European Commission's COIN Tool User Guide \cite{saisana_coin_2019} provides practical guidance for building composite indicators that are specifically designed for policy analysis. This guide serves as a useful resource for researchers and policymakers who aim to apply these metrics to evaluate and compare policy impacts across various regions or countries.

Drawing on these lessons, the Global AI Vibrancy Tool (GVT) applies best practices from the literature, which ensures a solid conceptual framework, transparent data handling, and thorough robustness checks. More details are provided in the \nameref{sec:methodology} Section.

There is also a well-established tradition of creating indices to track the technological progress of different nations. For example, the Technology Achievement Index (TAI), developed by \citet{desai_measuring_2002}, is a foundational framework for measuring cross-country technological advancement. The TAI evaluates countries according to several dimensions, including technology creation, diffusion, and human skill development. This index has set a foundation for more specialized tools designed to assess AI capabilities.

\citet{incekara_measuring_2017} developed TAI-16 from the original TAI, categorizing countries by their tech adoption and innovation. This index stresses how the dynamic pace of technological development requires frequently updated criteria. TAI-16 also shows how countries adapt to technological changes and measures AI readiness.

\citet{archibugi_technological_2009} offer an extensive analysis of synthetic indicators for measuring the technological capabilities of nations. Their work reviews various composite indicators developed by entities such as the European Commission, the World Economic Forum, and the World Bank, discussing their methodologies, assumptions, and consistency in results. These authors illustrate the importance of these indicators for public policy, corporate strategy, and economic studies, while addressing the challenges and limitations inherent in their use, such as potential oversimplification and the difficulty in capturing the full complexity of technological change.

\citet{shoham_toward_2017} was one of the first to propose systematically measuring and tracking the national progress and impact of Artificial Intelligence across various dimensions.  \citet{shoham_toward_2017} argues that there is a need for a multi-faceted country-level index of AI technologies, which includes factors such as investment, research output, and technological achievements. This proposed index set the ground for the AI Index \cite{maslej_ai_2024} initiative.

The Oxford Insights' AI Readiness Index \cite{hankins_2023_nodate} evaluates a nation's readiness to use AI in public service, taking into consideration elements such as innovation potential, data availability, infrastructure, and human capital. Furthermore, the AI Preparedness Index \cite{cazzaniga_gen-ai_2024} deals with a wider spectrum, including technological infrastructure, AI research communities, AI adoption by industry and government support. The Global AI Index by Tortoise Media \cite{cesareo_global_nodate} ranks countries based on research, development, talent, infrastructure, and operating environment.

The OECD AI Policy Observatory is another valuable resource \cite{noauthor_live_2024} which offers a comprehensive catalog of national AI strategies and provides insight into the strategic priorities and policy measures adopted by various countries. The Observatory collects data on AI policies, research investments, and other relevant factors from various countries. This data facilitates the assessment of different regulatory environments on AI development. This platform is important for tracking the AI-readiness of various nations. 

In conclusion, the related literature emphasizes the importance of composite indicators in providing structured and quantifiable assessments of complex phenomena such as AI development. However, it also cautions against oversimplification and stresses the need for continuous refinement of methodologies. These insights have informed the development of the GVT, which aims to deliver an authoritative and up-to-date measure of global AI vibrancy.

\subsection{Comparison with Other AI Indices and Tools}

Table \ref{tab: comparison_w_others} compares the GVT with other prominent AI indices and tools, summarizing the unique features, scope, indicators, and target audiences of each tool.

\begin{table}[h!]
\centering
\caption{Comparison of the Global AI Vibrancy Tool with Other AI Indices and Tools}
\small
\begin{adjustbox}{max width=\textwidth}
\begin{threeparttable}
\begin{tabular}{|>{\centering\arraybackslash}m{0.4cm}|>{\raggedright\arraybackslash}p{2.3cm}|>{\raggedright\arraybackslash}p{2.7cm}|>{\raggedright\arraybackslash}p{2.5cm}|>{\raggedright\arraybackslash}p{2.5cm}|>{\raggedright\arraybackslash}p{2.5cm}|>{\raggedright\arraybackslash}p{2.5cm}|}
\hline
& \textbf{Feature} & \textbf{Global AI Vibrancy Tool} & \textbf{Government AI Readiness Index} & \textbf{AI Preparedness Index} & \textbf{Global AI Index} & \textbf{OECD AI Policy Observatory} \\
\hline
\multirow{3}{*} & \textbf{Scope} & Cross-country comparisons of AI vibrancy & Government readiness for AI in public services & Readiness across strategic areas for AI adoption & Ranking of countries based on AI implementation, innovation, and investment & Comprehensive catalog of national AI strategies and policies \\
\cline{2-7}
{\rotatebox[origin=c]{90}{\textbf{Overview} \hspace{0.5cm}}} & \vspace{-0.6cm}\textbf{Focus Areas} & \begin{itemize}[leftmargin=*] \vspace{-0.6cm}\item Holistic view of AI development \item Detailed analysis of specific metrics over time\end{itemize} & \begin{itemize}[leftmargin=*] \vspace{-0.6cm} \item Government policy and implementation for public services \end{itemize} & \begin{itemize}[leftmargin=*] \vspace{-0.6cm} \item Comprehensive AI readiness \item Strategic areas\end{itemize} & \begin{itemize}[leftmargin=*] \vspace{-0.6cm} \item Practical implementation of AI, innovation, and investment \end{itemize} & \begin{itemize}[leftmargin=*] \vspace{-0.6cm} \item AI policy impacts \item Strategic priorities \item Policy measures\end{itemize} \\
\cline{2-7}
& \textbf{Target Audience} & Policymakers, industry leaders, researchers, general public & Policymakers, government officials & Policymakers, industry leaders, researchers & Policymakers, industry leaders, researchers, general public & Policymakers, industry leaders, researchers, general public \\
\hline
\multirow{4}{*} & \textbf{No. of Indicators}\tnote{*} & \multicolumn{1}{c|}{42} &  \multicolumn{1}{c|}{39} &  \multicolumn{1}{c|}{29} &  \multicolumn{1}{c|}{122} &  \multicolumn{1}{c|}{118} \\
\cline{2-7}
& \textbf{No. of Dimensions} &  \multicolumn{1}{c|}{8} &  \multicolumn{1}{c|}{10} &  \multicolumn{1}{c|}{4} &  \multicolumn{1}{c|}{3} &  \multicolumn{1}{c|}{11} \\
\cline{2-7}
& \textbf{No. Countries} &  \multicolumn{1}{c|}{36} &  \multicolumn{1}{c|}{193} &  \multicolumn{1}{c|}{174} &  \multicolumn{1}{c|}{83} &  \multicolumn{1}{c|}{60} \\
\cline{2-7}
& \textbf{Overarching Index} &  \multicolumn{1}{c|}{\checkmark} & \multicolumn{1}{c|}{\checkmark} & \multicolumn{1}{c|}{\checkmark} & \multicolumn{1}{c|}{\checkmark} & \multicolumn{1}{c|}{\ding{55}} \\
\cline{2-7}
& \textbf{Absolute vs. Relative Rankings} &  \multicolumn{1}{c|}{\checkmark} & \multicolumn{1}{c|}{\ding{55}} & \multicolumn{1}{c|}{\ding{55}} & \multicolumn{1}{c|}{\checkmark} & \multicolumn{1}{c|}{\ding{55}} \\
\cline{2-7}
& \textbf{Publicly Available Data} &  \multicolumn{1}{c|}{\checkmark} & \multicolumn{1}{c|}{\ding{55}} & \multicolumn{1}{c|}{\ding{55}} & \multicolumn{1}{c|}{\ding{55}} & \multicolumn{1}{c|}{\checkmark} \\
\cline{2-7}
{\rotatebox[origin=c]{90}{\textbf{Attributes}\hspace{1.2cm}}} & \vspace{-0.8cm} \textbf{Data Presentation} & \begin{itemize}[leftmargin=*] \vspace{-0.8cm} \item High number of
AI-related
indicators \item 
Adjustable
weights \item Downloadable\end{itemize} & \begin{itemize}[leftmargin=*] \vspace{-0.8cm} \item Focused on
public sector
readiness \item Static rankings and scores \end{itemize} & \begin{itemize}[leftmargin=*] \vspace{-0.8cm} \item Broad strategic
coverage \item Static scores \end{itemize} & \begin{itemize}[leftmargin=*] \vspace{-0.8cm} \item Broad AI and non-AI indicators \item Static rankings and scores \end{itemize} & \begin{itemize}[leftmargin=*] \vspace{-0.8cm} \item Broad
policy catalog \item Some live data feeds \item Downloadable\end{itemize} \\
\cline{2-7}
& \textbf{User Experience} & \begin{itemize}[leftmargin=*] \vspace{-0.25cm} \item User-friendly navigation \item Detailed interface \item Dynamic and interactive features\end{itemize} & \begin{itemize}[leftmargin=*] \vspace{-0.25cm} \item Clear navigation \item Clean interface \item Simple features\end{itemize} & \begin{itemize}[leftmargin=*] \vspace{-0.25cm} \item Simple navigation \item Clean interface \item Simple features\end{itemize} & \begin{itemize}[leftmargin=*] \vspace{-0.25cm} \item Easy-to-use navigation \item User-friendly interface \item Informative features\end{itemize} & \begin{itemize}[leftmargin=*] \vspace{-0.25cm} \item Easy-to-use navigation \item Organized interface \item Dynamic and interactive features\end{itemize} \\
\cline{2-7}
& \textbf{Customization} & \multicolumn{1}{c|}{\checkmark} & \multicolumn{1}{c|}{\ding{55}} & \multicolumn{1}{c|}{\ding{55}} & \multicolumn{1}{c|}{\checkmark} & \multicolumn{1}{c|}{\checkmark} \\
\hline
\end{tabular}
\begin{tablenotes}
\footnotesize
\item[*] Differences in indicator counts across sources may result from varying counting methods. Some sources count relative indicators (e.g., per capita or per GDP) as separate from their absolute counterparts, while others do not.
\end{tablenotes}
\end{threeparttable}
\end{adjustbox}
\label{tab: comparison_w_others}
\end{table}

The new GVT serves the purpose of benchmarking national progress in AI and fills some of the gaps that currently exist in the AI national vibrancy tracking landscape. Unlike the Government AI Readiness Index \cite{hankins_2023_nodate}, which primarily evaluates AI readiness in public services, or the AI Preparedness Index \cite{cazzaniga_gen-ai_2024}, which focuses on strategic areas for AI adoption, the GVT employs diverse indicators (for example in research and development, the economy or responsible AI) that are organized into distinct pillars housed under an overarching index. This structure facilitates broader possibilities of analysis and creates a tool that can be more widely and flexibly used by the broader AI community.

The tool's selectivity means that it presents reliable data on AI-related activities, mainly focusing on metrics specifically related to AI rather than broader technological metrics like ``total amount of public spending in R\&D'' (as do other tools like the Global AI Index \cite{cesareo_global_nodate}). Moreover, the GVT excels in data presentation, offering interactive, customizable, and downloadable visualizations, which improves user engagement and comprehension. This feature contrasts with the more static presentations found in other indices. While the OECD AI Policy Observatory \cite{noauthor_live_2024} is rich in data and interactive visualizations, it lacks an index or ranking system.

Moreover, a key differentiator of the GVT is its commitment to openness: all data is public and users are allowed to flexibly adjust pillar and indicator weights. This flexibility improves the tool's applicability across different use cases and allows users to bring their own perspectives to the question of which particular AI pillars (for instance research and development versus policy and governance) matter most in judging a nation's AI vibrancy.

The GVT is built to respond more effectively to the fast-evolving AI landscape compared to static indices. While some tools may have strong coverage in specific areas, they often lack the flexibility and AI-related focus offered by this tool.

\newpage

\section{Conceptual Framework}
\label{sec:conceptual-framework}

The primary objective of the Global AI Vibrancy Tool (GVT) is to facilitate cross-country comparisons of AI vibrancy in the field of AI. AI vibrancy can be defined as the level of activity, development, and impact of AI technologies within a country. This assessment provides a comprehensive understanding of the progress different countries are making in AI thereby highlighting strengths and areas for potential improvement.

The GVT captures country-level AI vibrancy through several key dimensions, each representing a critical aspect of AI development. The dimensions, henceforth referred to as pillars, include: Research and Development (R\&D), Responsible AI, Economy, Education, Diversity, Policy and Governance, Public Opinion, and Infrastructure. Each pillar is represented by a group of indicators that serve as proxies to capture the underlying concept.

\vspace{12pt}
\noindent

\begin{enumerate}[label=\arabic*.]
    \item 
    \textbf{Research and Development (R\&D)}
    
    \noindent
    R\&D is the foundation of AI advancement, driving the creation of new algorithms, models, and technologies that in turn foster AI innovation. According to \citet{furman_determinants_2002}, national innovative capacity, defined as the long-term ability to produce and commercialize innovative technology, is significantly influenced by R\&D efforts. Measures of innovative output, such as patenting activities and journal publications, serve as key indicators of this capacity.
    
    \item 
    \textbf{Responsible AI}
    
    \noindent
    Building AI systems that adhere to ethical standards is important for gaining public trust and preventing harm. Responsible AI covers several dimensions including data governance, explainability, fairness, privacy, security and safety, and transparency among others \cite{maslej_ai_2024}. The volume of conference submissions related to these aspects affiliated with a country can serve as a quantitative measure of ongoing research and discussion on responsible AI practices.\footnote{As this is a relatively new field, currently the AI Index has relatively few Responsible AI indicators. The AI Index team is working to include more Responsible AI indicators for future iterations of the tool. The AI Index recognizes that many factors contribute to a country’s Responsible AI vibrancy—such as the extent to which businesses operationalize Responsible AI—beyond those currently captured in the tool. While the metrics featured, such as conference submissions, are not intended to fully represent a country’s Responsible AI vibrancy, the Index team included a Responsible AI pillar for two key reasons: first, to underscore the importance of Responsible AI within the broader AI ecosystem, and second, to encourage the AI research community to develop and contribute additional Responsible AI metrics in the future. To address the data limitations of the Responsible AI pillar, it has been assigned a lower weight, as detailed in the methodological section.}

    \item 
    \textbf{Economy}
    
    \noindent
    The economic landscape surrounding AI is a key factor influencing how AI is developed and deployed. In their work on the economic implications of AI, \citet{agrawal_economics_2019} highlight factors such as research investment, infrastructure, applications, and  labor market conditions, all of which play an important role in guiding AI innovation. Trends in investment volumes and job markets can capture some of these aspects.
    
    \item
    \textbf{Education}
    
    \noindent
    Education is essential for preparing a skilled AI workforce. As \citet{pedro_artificial_2019} state, preparing the future workforce for AI involves more than just adopting advanced technologies. It necessitates a shift in curricula to emphasize ``AI competencies.'' Tracking the growth of AI-related degree programs can provide insights into the effectiveness of these efforts.

    \item
    \textbf{Diversity}
    
    \noindent
    Diversity in AI development means accommodating and working with a wide range of perspectives which can potentially reduce biases in AI systems. It includes gender, ethnicity, and socioeconomic diversity within the AI community. Integrating diversity and inclusion principles throughout the AI lifecycle is important for creating fair and transparent AI technologies \cite{zowghi_diversity_2023}. Diverse teams bring varied experiences and viewpoints, which can be important in identifying and mitigating biases. One way to capture diversity is by measuring AI skills across genders.\footnote{While there are currently limited diversity metrics, the AI Index is working to include more in future iterations of the tool.}
    
    \item
    \textbf{Policy and Governance}
    
    \noindent
    Policy and governance frameworks set the base for AI ecosystems, influencing everything from innovation and ethical standards to investments and education. National AI strategies, which are policy plans created by governments to guide the development and deployment of AI within their country, often integrate efforts from both the public and private sectors, emphasizing ethics over rigid rule-based systems \cite{radu_steering_2021}. Some nations opt for self-regulation and market-based approaches, while others stress public responsibility and ethical considerations \cite{djeffal_role_2022}. Tracking the number of AI-related legislation, strategies and legislative mentions can provide valuable insight into how different countries approach AI governance and regulation.

    \item
    \textbf{Public Opinion}
    
    \noindent
    Public perception of AI influences its adoption and development. Understanding public opinion helps address concerns and improve AI literacy. The framing of AI in media affects public sentiment and, consequently, regulatory decisions and technology acceptance. Comprehending public opinion, including analyzing media conversations in terms of volume and sentiment, can reveal important insights on the level of public support for AI development within a nation. \cite{sartori_minding_2023}.
    \item
    \textbf{Infrastructure}
    
    \noindent
    Robust infrastructure is a critical prerequisite for advancing AI research and deployment, with significant variations observed across countries. Infrastructure includes computational resources, data availability, and network connectivity. \citet{amodei_ai_2018} demonstrate that the computational power used in the largest AI training runs has been increasing exponentially. Countries with superior data centers and cloud infrastructure gain a significant advantage in conducting large-scale AI experiments. Moreover, the quality and quantity of available data sets vary widely between nations, directly impacting their ability to develop accurate and generalizable AI models. Network connectivity further amplifies these disparities. Nations that possess faster and more reliable internet connections can more quickly process real-time data \cite{hankins_2023_nodate}. The number of supercomputers, compute capacity, and high-speed internet can capture some infrastructure-related concepts essential for AI advancements.
    
\end{enumerate}

These dimensions can be combined to provide a comprehensive picture of a nation's AI vibrancy. High levels of R\&D activity lead to research innovations, which can be translated into commercial success and widespread adoption in supportive economic environments. Responsible AI practices ensure that AI advancements are ethical and sustainable, which can lead to greater public acceptance. Higher rates of AI trust can in turn encourage regulatory and commercial environments that are more friendly towards AI.

Robust education ecosystems can provide a continuous supply of skilled professionals that can drive AI forward, while diversity within the AI community improves creativity and reduces biases. AI solutions that are less biased can be more effective and equitable. Robust policy and governance frameworks provide the necessary support and direction for AI development. On the one hand, these policies can safeguard the deployment of AI further engendering public trust. On the other hand they can create incentives for better commercial AI adoption and a more robust AI research and development ecosystem.

Public opinion shapes the adoption and integration of AI technologies into society, while strong infrastructure supports the efficient development and deployment of these technologies. Together, these dimensions have a synergistic effect: improvements in one dimension can drive progress in another, ultimately leading to an overall increase in a nation's AI vibrancy.

\subsection{Sub-Indices for Granular Comparison}
\label{subsec:sub-indices}

To facilitate the comparative analysis of countries at a granular level, we introduce sub-indices within the GVT. These sub-indices provide a more nuanced perspective on particular aspects of AI development and are intended to enable a more precise evaluation of specific dimensions. While the overarching index offers a comprehensive overview of a country's aggregate AI development, the sub-indices facilitate a more in-depth examination of particular domains.

\vspace{12pt}
\noindent

\begin{enumerate}[label=\arabic*.]
    \item 
    \textbf{Innovation Index}
    
    \noindent 
    This index measures a country's innovation potential by assessing its research and development (R\&D) activities, academic output, technological advancements, intellectual property generation, and supporting technological infrastructure. It highlights a country's capacity to produce new knowledge, innovate, and contribute to global AI advancements. The index offers insights into which countries are leading in AI development and have the necessary infrastructure to support future progress. 

    \item
    \textbf{Economic Competitiveness Index}

    \noindent
    This index measures the economic strength and market dynamism of a country in the AI sector by analyzing investment flows, talent concentration, and job creation. It captures how countries are leading in integrating AI into their economies, enhancing their competitive edge, and creating robust ecosystems for AI-driven growth.

    \item
    \textbf{Policy, Governance, and Public Engagement Index}
    
     \noindent
    This index evaluates the level of activity related to AI-related policies, legislative actions, and the broader public discourse surrounding AI. It identifies which countries are leaders in creating an enabling policy environment for AI and how public opinion is shaping the adoption and acceptance of AI technologies. Countries with a national strategy on AI and positive public sentiment are better positioned to leverage AI for national development, while those lagging may face challenges in governance, public trust, and AI deployment.

\end{enumerate}

\vspace{12pt}

\section{Methodology}

\label{sec:methodology}
\subsection{Data Collection}
\label{sec:data-collection}
Our data collection strategy works to achieve comprehensive and reliable data. We gather information from sources like CSET, QUID, GitHub, and LinkedIn, as well as our prepared datasets.

Maintaining data integrity is a top priority. Our approach includes implementing rigorous validation processes to cross-check data from multiple sources, in order to maintain consistency and accuracy. We periodically update our database to incorporate the most recent and comprehensive data, in order to guarantee that our tool reflects current trends and developments.

The selection of data to be included in our tool is based on the following criteria:
\begin{itemize}
  \item \textbf{Relevance}: The data must be directly relevant to the dimensions of AI vibrancy we aim to measure, such as R\&D, Economy, Education, etc.
  
  \item \textbf{Significance}: Indicators should have a significant impact on AI vibrancy, providing meaningful insights into a country's AI capabilities and development.
  
  \item \textbf{Accuracy}: Data must be accurate and reliable, sourced from reputable providers or validated through rigorous self-collection methods.
  
  \item \textbf{Coverage}: Indicators should have good geographical and temporal coverage, allowing for comprehensive comparisons across different countries and over time.
  
  \item \textbf{Traceability}: Data should be easy to track and updated annually, ensuring that our tool can provide up-to-date and consistent evaluations.
\end{itemize}

Despite our attentive data collection strategy, several challenges arise. Some indicators have limited coverage across countries or over time. For example, we considered including the percentage of businesses using AI technology as an indicator. However,  available surveys containing this data lack sufficient country coverage. Similarly, we aimed to include public opinion data on AI from surveys. Unfortunately, these surveys are not always repeated annually, and the questions change over time, leading to inconsistencies in the data. To maintain the global applicability of the tool, we prioritize indicators with the broadest possible country and time coverage.

Data quality can vary by country and over time. For instance, we attempted to gather data on the number of graduates in various countries by contacting official statistical offices. However, for some countries, the data was either incomplete, not available at a more detailed level, or no response was received. Additionally, changes in data collection methodologies over time can affect the consistency of some indicators. We carefully document and adjust for any methodological changes to maintain the integrity of our longitudinal analyses.

The 42 included indicators across 8 different pillars are described in Table \ref{tab:ai_indicators}.\footnote{See the list of countries included in the Appendix, Section \ref{sec:appendix-countries}. Data coverage by country and by indicator for each year is provided in the Appendix, Section \ref{sec:appendix-coverage}. Definitions of each indicator are provided in the Appendix, Section \ref{sec:appendix-indicators-descriptions}.} \\

\begin{table}[H]
\centering
\footnotesize
\caption{List of Indicators by Pillar.}
\begin{adjustbox}{max width=\textwidth}
\begin{tabular}{|>{\raggedright\arraybackslash}p{4cm}|>{\raggedright\arraybackslash}p{6cm}|>{\raggedright\arraybackslash}p{3cm}|}
\hline
\textbf{Pillar} & \textbf{Indicator} & \textbf{Data Source} \\ \hline
\cellcolor{researchdevelopment} \textcolor{white}{Research and Development} & AI Journal Publications & Center for Security and Emerging Technology \\ \hline
\cellcolor{researchdevelopment} \textcolor{white}{Research and Development} & AI Journal Citations & Center for Security and Emerging Technology \\ \hline
\cellcolor{researchdevelopment} \textcolor{white}{Research and Development} & AI Conference Publications & Center for Security and Emerging Technology \\ \hline
\cellcolor{researchdevelopment} \textcolor{white}{Research and Development} & AI Conference Citations & Center for Security and Emerging Technology \\ \hline
\cellcolor{researchdevelopment} \textcolor{white}{Research and Development} & AI Patent Grants & Center for Security and Emerging Technology \\ \hline
\cellcolor{researchdevelopment} \textcolor{white}{Research and Development} & Notable Machine Learning Models & Epoch AI \cite{epoch_ai_data_2024} \\ \hline
\cellcolor{researchdevelopment} \textcolor{white}{Research and Development} & Academia-Industry Model Production Concentration\tablefootnote{See calculation details in the Appendix, Section \ref{sec:appendix-model-production}.}
 & Epoch AI \cite{epoch_ai_data_2024}; AI Index \\ \hline
\cellcolor{researchdevelopment} \textcolor{white}{Research and Development} & Foundation Models & Ecosystem Graphs \cite{bommasani_ecosystem_2023}; AI Index \\ \hline
\cellcolor{researchdevelopment} \textcolor{white}{Research and Development} & Foundation Models Datasets & Ecosystem Graphs \cite{bommasani_ecosystem_2023}; AI Index \\ \hline
\cellcolor{researchdevelopment} \textcolor{white}{Research and Development} & Foundation Models Applications & Ecosystem Graphs \cite{bommasani_ecosystem_2023}; AI Index \\ \hline
\cellcolor{researchdevelopment} \textcolor{white}{Research and Development} & Open Access Foundation Models & Ecosystem Graphs \cite{bommasani_ecosystem_2023}; AI Index \\ \hline
\cellcolor{researchdevelopment} \textcolor{white}{Research and Development} & AI GitHub Projects & GitHub \\ \hline
\cellcolor{researchdevelopment} \textcolor{white}{Research and Development} & AI GitHub Projects Stars & GitHub \\ \hline
\cellcolor{responsibleai} \textcolor{white} {Responsible AI} & FAccT Conference Submissions on RAI Topics & AI Index \\ \hline
\cellcolor{responsibleai} \textcolor{white} {Responsible AI} & NeurIPS Conference Submissions on RAI Topics & AI Index \\ \hline
\cellcolor{responsibleai} \textcolor{white} {Responsible AI} & ICML Conference Submissions on RAI Topics & AI Index \\ \hline
\cellcolor{responsibleai} \textcolor{white} {Responsible AI} & ICLR Conference Submissions on RAI Topics & AI Index \\ \hline
\cellcolor{responsibleai} \textcolor{white} {Responsible AI} & AIES Conference Submissions on RAI Topics & AI Index \\ \hline
\cellcolor{responsibleai} \textcolor{white} {Responsible AI} & AAAI Conference Submissions on RAI Topics & AI Index \\ \hline
\cellcolor{economy} \textcolor{white} {Economy} & Total AI Private Investment & QUID \\ \hline
\cellcolor{economy} \textcolor{white} {Economy} & Total AI Merger/Acquisition Investment & QUID \\ \hline
\cellcolor{economy} \textcolor{white} {Economy} & Total AI Minority Stake Investment & QUID \\ \hline
\cellcolor{economy} \textcolor{white} {Economy} & Total AI Public Offering Investment  & QUID \\ \hline
\cellcolor{economy} \textcolor{white} {Economy} & Newly Funded AI Companies & QUID \\ \hline
\cellcolor{economy} \textcolor{white} {Economy} & AI Hiring Rate YoY Ratio & LinkedIn \\ \hline
\cellcolor{economy} \textcolor{white} {Economy} & Relative AI Skill Penetration & LinkedIn \\ \hline
\cellcolor{economy} \textcolor{white} {Economy} & AI Talent Concentration & LinkedIn \\ \hline
\cellcolor{economy} \textcolor{white} {Economy} & AI Job Postings (\% of Total) & Lightcast \\ \hline
\cellcolor{economy} \textcolor{white} {Economy} & Net Migration Flow of AI Skills & LinkedIn \\ \hline
\cellcolor{education} \textcolor{white} {Education} & AI Study Programs in English & Studyportals \\ \hline
\cellcolor{education} \textcolor{white} {Education} & AI Study Programs in English Penetration & Studyportals \\ \hline
\cellcolor{diversity} \textcolor{white} {Diversity} & AI Talent Concentration Gender Equality Index\tablefootnote{See calculation details in the Appendix, Section \ref{sec:appendix-ai-talent-concentration-gender}.} & LinkedIn; AI Index \\ \hline
\cellcolor{policygovernce} \textcolor{white} {Policy and Governance} & National AI Strategy Presence & AI Index \\ \hline
\cellcolor{policygovernce} \textcolor{white} {Policy and Governance} & AI Legislation Passed & AI Index \\ \hline
\cellcolor{policygovernce} \textcolor{white} {Policy and Governance} & AI Mentions in Legislative Proceedings & AI Index \\ \hline
\cellcolor{publicopinion} \textcolor{white} {Public Opinion} & Social Media Share of Voice on AI & QUID \\ \hline
\cellcolor{publicopinion} \textcolor{white} {Public Opinion} & AI Social Media Posts & QUID \\ \hline
\cellcolor{publicopinion} \textcolor{white} {Public Opinion} & AI-Related Social Media Conversations Net Sentiment & QUID \\ \hline
\cellcolor{infrastructure} \textcolor{white} {Infrastructure} & Parts Semiconductor Devices Exports & BACI \cite{gaulier_baci_2010} \\ \hline
\cellcolor{infrastructure} \textcolor{white} {Infrastructure} & Supercomputers & Top500 \cite{noauthor_top500_2024} \\ \hline
\cellcolor{infrastructure} \textcolor{white} {Infrastructure}  & Compute Capacity (Rmax) & Top500 \cite{noauthor_top500_2024} \\ \hline
\cellcolor{infrastructure} \textcolor{white} {Infrastructure}  & Internet Speed & Ookla \cite{noauthor_ooklas_2024} \\ \hline
\end{tabular}
\label{tab:ai_indicators}
\end{adjustbox}
\end{table}

\subsection{AI Vibrancy Index Construction}
The index construction process involves several key steps:\footnote{For more details on standard composite indicator construction practices, see \citet{oecd_handbook_2008}.} (1) normalizing the data, (2) calculating the pillar scores, and (3) aggregating these scores to form the overall AI vibrancy index. Below is a detailed explanation of each step.

\subsubsection*{Normalization}
To ensure comparability across different scales and units of measurement, we use min-max normalization for all indicators. This method scales the values within the [0, 100] range. The normalized value for the ${i}^{th}$ indicator is calculated using the formula:

\begin{equation}
x_{ijk} = \frac{x_{ijk}^{\text{raw}} - x_{ij, \text{min}}}{x_{ij, \text{max}} - x_{ij, \text{min}}} \cdot 100
\end{equation}

\noindent where:
\begin{itemize}
    \item $x_{ijk}^{\text{raw}}$ is the original value of the $i^{th}$ indicator for pillar $j$ and country $k$.
    \item $x_{ij, \text{min}}$ and $x_{ij, \text{max}}$ are the minimum and maximum values of indicator $i$ in pillar $j$ across all countries.
\end{itemize}

\subsubsection*{Pillar Score Calculation}

Let $p_{jk}$ denote the score of pillar $j$ for country $k$. This score is computed as the weighted average of various indicators associated with the pillar and the country. Specifically, the score is given by the formula:

\begin{equation}
p_{jk} = \frac{\sum_{i=1}^{N_j} (w_{ij} \cdot x_{ijk})}{\sum_{i=1}^{N_j} w_{ij}}
\end{equation}

where:
\begin{itemize}
    \item $N_j$ is the number of indicators for pillar $j$.
    \item $w_{ij}$ is the weight assigned to the $i^{th}$ indicator for pillar $j$, ranging from 0 to 10, with higher values indicating greater importance of the indicator. We assume that these weights are non-negative and their sum is non-zero.
    \item $x_{ijk}$ is the normalized score of the $i^{th}$ indicator for pillar $j$ and country $k$.
\end{itemize}

\subsubsection*{AI Vibrancy Index Calculation}

The AI vibrancy index for a country $k$, denoted as $V_{k}$, is calculated as the weighted average of the scores of all the pillars as follows:

\begin{equation}
V_{k} = \frac{\sum_{j=1}^{M} (W_j \cdot p_{jk})}{\sum_{j=1}^{M} W_j}
\end{equation}

where:
\begin{itemize}
    \item $M$ is the total number of pillars.
    \item $W_j$ is the weight assigned to pillar $j$, ranging from 0 to 10, with higher values indicating greater importance of the pillar. We assume that these weights are non-negative and their sum is non-zero.
    \item $p_{jk}$ is the score of pillar $j$ for country $k$, as calculated above.
\end{itemize}

This formulation allows for a comprehensive assessment of a country's vibrancy, taking into account a range of factors across different dimensions.

\subsubsection*{Ranking Countries}
After calculating the AI vibrancy score for each country, we rank the countries for each year by their resulting weighted index scores. This ranking offers an annual comparison of AI vibrancy across different countries, whether in absolute or per capita terms.

\subsubsection*{Time Dimension}
The AI Vibrancy index computation detailed above is calculated for data aggregated over a calendar year. The tool then permits tracking the evolution of the absolute scores as well as the relative rankings across time.\footnote{Note that the interpretation of the index over time may be influenced by the availability of data for specific indicators. In some cases, certain indicators may not be available for all years or countries, potentially affecting the comparability of the index across time periods. The index values for these periods should be interpreted with caution, as they may reflect variations in data coverage. For detailed information on data availability, refer to the accompanying data coverage tables in the methodology section of the tool webpage.}

\vspace{12pt}

This process is applied to each sub-index: Innovation, Economic Competitiveness, and Policy, Governance, and Public Engagement. We aggregate only the relevant indicators listed in the corresponding tables provided in the Appendix, Section \ref{sec:appendix-subindices-indicators}, ensuring each sub-index provides a focused analysis of a specific AI vibrancy dimension.

\subsection{Implementation}
\subsubsection{Handling Missing Values}
Missing values have been imputed with the median of each indicator across all countries for each year. If an indicator is missing for all countries in a specific year, it is excluded from the calculation and its weight is proportionally redistributed among the other available indicators. As a result, the total weight assigned to the remaining indicators sums to the original total, thus maintaining the overall balance of importance among indicators.

This approach ensures that the data remains as complete as possible, minimizing the loss of information due to missing values. By imputing with the median, we preserve the integrity of the data while reducing potential biases that might arise from more complex imputation methods. Excluding indicators entirely when data is unavailable for all countries in a year prevents the introduction of inaccuracies and ensures that the analysis reflects the best available data.

However, this methodology has its trade-offs, including the potential exclusion of key indicators that could impact the overall assessment. Additionally, variability in data availability across years may complicate temporal comparisons. Future enhancements could include a feature for users to run sensitivity analyses, allowing them to assess the impact of different imputation strategies on the ranking. Promoting comprehensive and consistent data reporting across more countries will further enhance the tool’s accuracy and reliability over time.

\subsubsection{Weights Selection}
The inclusion of all indicators in the GVT is based on the criteria defined in Section \ref{sec:data-collection}. However, in assessing a country's level of AI vibrancy, some pillars and indicators are admittedly more relevant than others. We set the default weights of the GVT using an expert budget allocation approach, where four senior AI Index team members individually allocated weights according to their relative perceived relevance, while also considering specific features of some indicators, such as coverage.\footnote{Alternative approaches exist, such as Principal Component Analysis (PCA), which statistically transforms correlated variables into a set of uncorrelated variables and derives weights based on the variance each component explains. However, this method can reduce the interpretability of the resulting metrics. A hybrid method involves using regression coefficients, where weights are determined through regression analysis guided by a theoretical model. Another common approach is to assign equal weights, which may oversimplify and introduce potential bias. See \citet{oecd_handbook_2008} for a comparison of different weighting methods.} For instance certain pillars such as Education and Responsible AI were deemed to be relatively important by the ``expert group'' as components of a nation's AI vibrancy. However, because the data for these pillars was limited in its international coverage (certain nations were better represented then others), these pillars were assigned a lower weight.\footnote{For more details on how weights were attributed, refer to the default weights selected for pillars and indicators in Appendix, Section \ref{sec:appendix-weighting}.}\footnote{The sub-indices use the same default weights as those applied in calculating the AI vibrancy index.} The median weight was then selected for each pillar and indicator. We recognize that this choice impacts the AI Vibrancy Index. The tool’s interface allows users to adjust the weights using interactive sliders, with values ranging from 0 to 10, where 0 represents lowest importance and 10 represents highest importance. This feature improves the tool’s adaptability, enabling users to customize the assessment according to their own perspective on the relative importance of pillars and indicators.

\section{Tool Interface Overview}
In this section we present an overview of the tool interface, highlighting its components and functionalities. The interface is designed to be user-friendly and interactive. Users can explore and customize views according to their preferences.

\subsection{Main Components}

\subsubsection{Global and National AI Vibrancy Rankings}
This section allows users to toggle between global and national rankings, providing a comparative view of AI vibrancy across different countries. The ``View Type'' component of the tool interface allows users to choose how to visualize the data that best suits their needs. The available view types are:

\begin{itemize}
    \item \textbf{Bar}: The bar chart (Figure \ref{fig: bar_view}) view presents the AI vibrancy ranking of different countries for a selected year. This view is particularly useful in comparing where each country stands in the ranking and which countries are leading or lagging in specific pillars.

    \begin{figure}[h!]
        \centering
        \includegraphics[width=\textwidth]{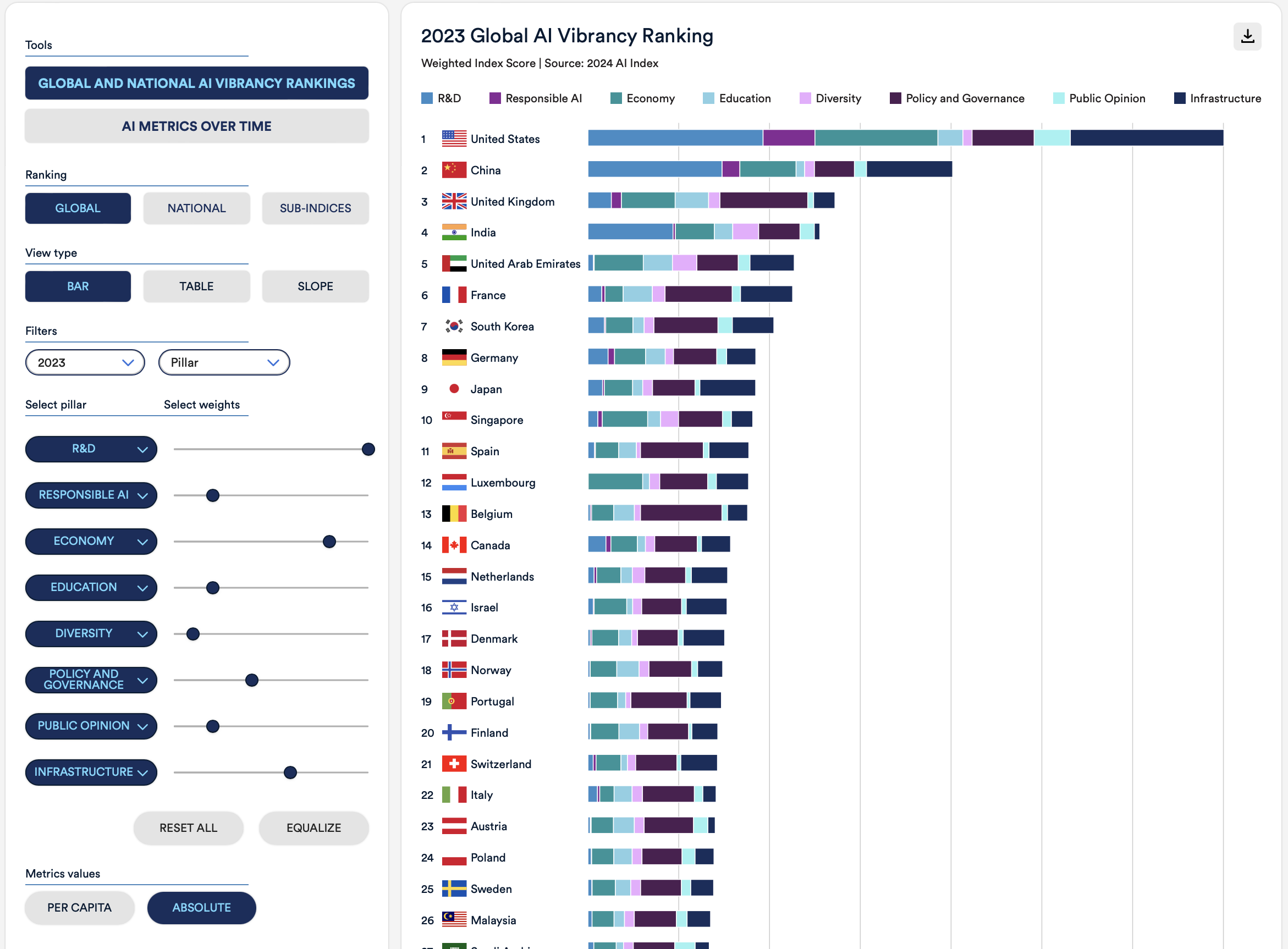}
        \caption{Global AI Vibrancy Ranking: Bar View.}
        \label{fig: bar_view}
    \end{figure}
    
    \item \textbf{Table}: The table view (Figure \ref{fig: table_view}) allows for comparisons of AI vibrancy across up to four selected countries simultaneously for a selected year. It includes information on each indicator within corresponding pillars, such as raw data values, normalized indicator scores, each country's overall rank, and the contribution of each indicator relative to other countries.

    \begin{figure}[h!]
        \centering
        \includegraphics[width=\textwidth]{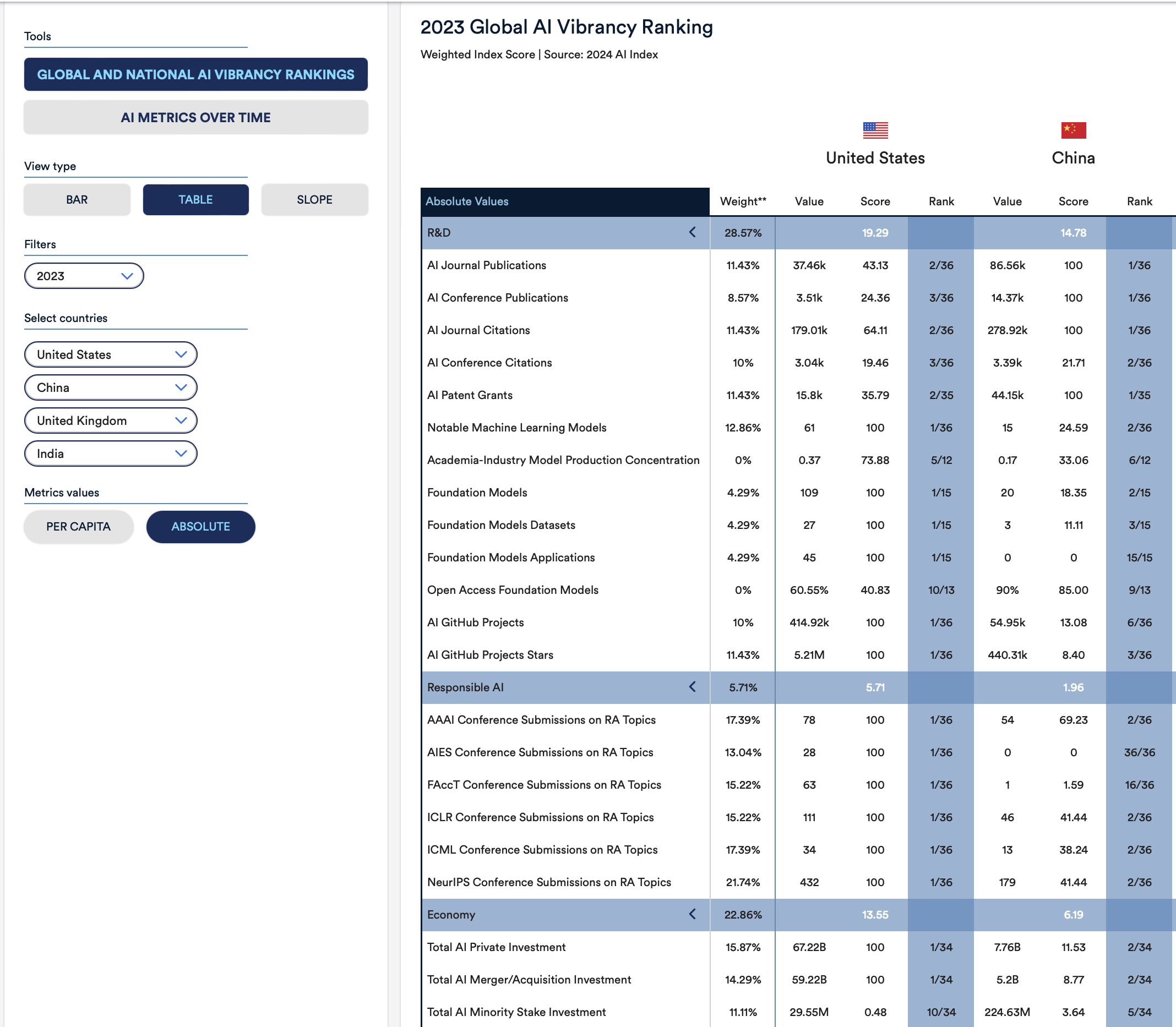}
        \caption{Global AI Vibrancy Ranking: Table View.}
        \label{fig: table_view}
    \end{figure}
    
    \item \textbf{Slope}: The slope view (Figure \ref{fig: slope_view}) displays changes in the global AI vibrancy ranking over time, showing the trajectory of each country.

    \begin{figure}[h!]
        \centering
        \includegraphics[width=\textwidth]{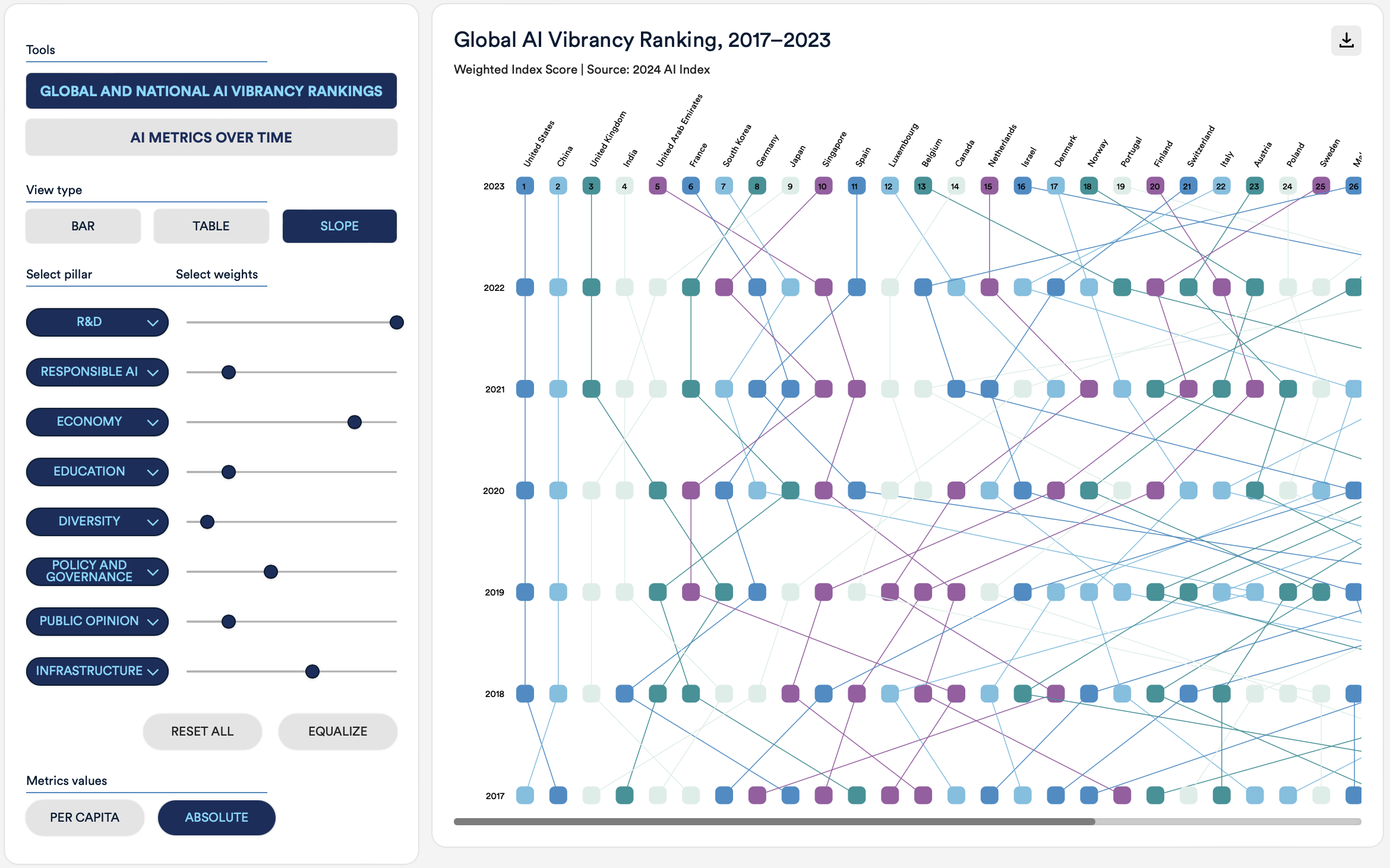}
        \caption{Global AI Vibrancy Ranking: Slope View.}
        \label{fig: slope_view}
    \end{figure}
    
\end{itemize}

\clearpage

\begin{figure}[H]
    \centering
    \includegraphics[width=0.9\linewidth]{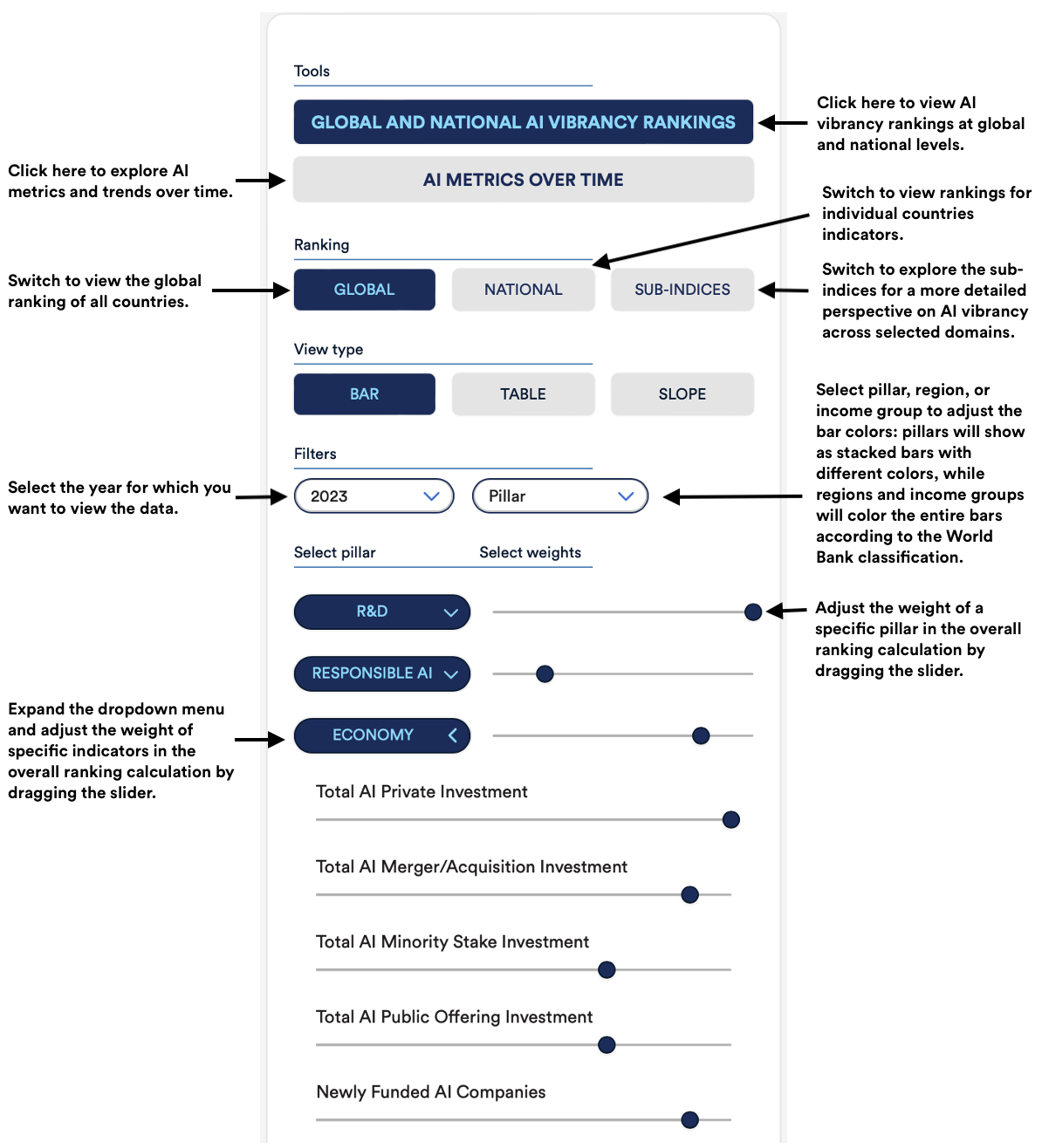}
    \caption{Interactive Filters and Customization Interface.}
    \label{fig: filters_customizations}
\end{figure}

The tool interface includes various filters and customization options (Figure \ref{fig: filters_customizations}) to enhance user experience and data analysis. Users can select the year for which they want to view the data, and toggle between per capita or absolute values. Additionally, controls allow users to color each bar according to pillar, region, and income group as per the World Bank country classification \cite{noauthor_wdi_nodate}.

\newpage

\subsubsection{AI Metrics Over Time}
This section provides users with a dynamic way to analyze changes in metrics included in the AI vibrancy index over time. Users can choose a country from a drop-down menu and select a metric from a specific pillar, allowing for a more focused analysis of a particular area of interest.

The tool offers two chart types for visualization:

\begin{itemize}
    \item \textbf{Bar}: This option shows AI metrics in a bar chart format (Figure \ref{fig: bar_view_metrics}), making it easy to compare values over different years.

    \begin{figure}[H]
        \centering
        \includegraphics[width=1\linewidth]{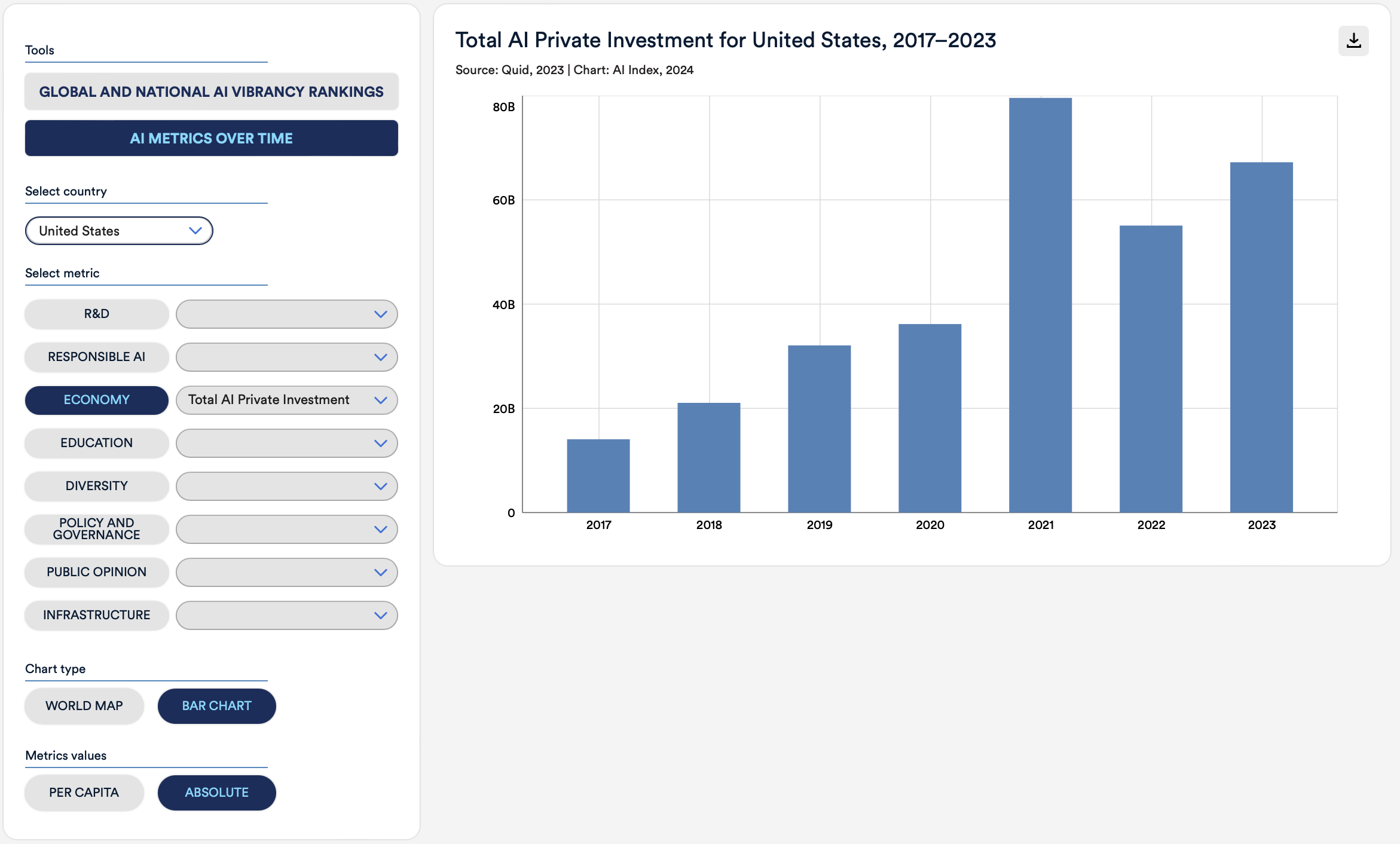}
        \caption{AI Metrics Over Time: Bar View.}
        \label{fig: bar_view_metrics}
    \end{figure}
    
    \item \textbf{World Map}: This option provides a geographical visualization of AI Metrics, showing the distribution and intensity of AI-related activities across countries (fig. \ref{fig: world_map_metrics}).
    
    \begin{figure}[H]
        \centering
        \includegraphics[width=1\linewidth]{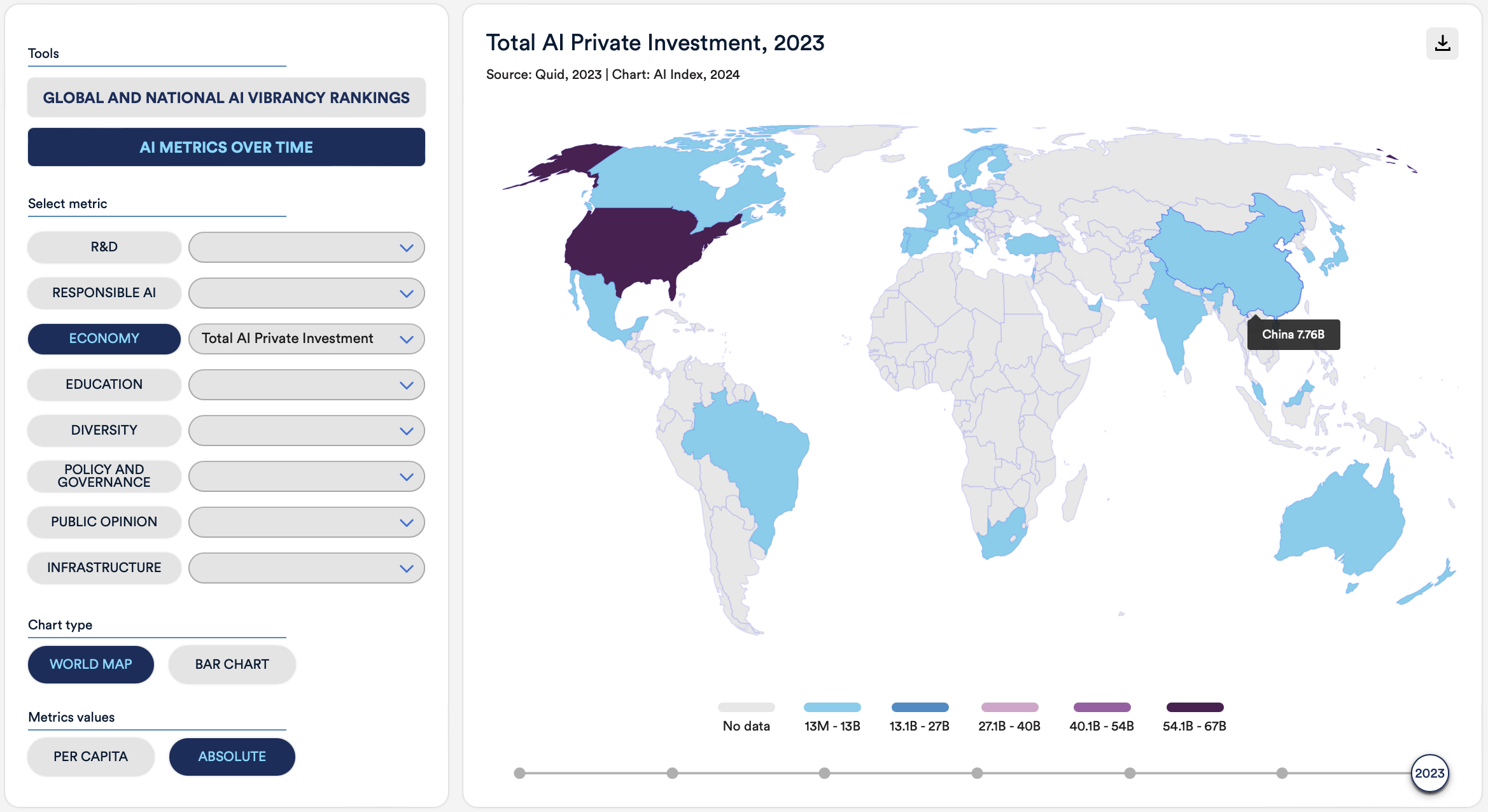}
        \caption{AI Metrics Over Time: World Map View.}
        \label{fig: world_map_metrics}
    \end{figure}
\end{itemize}

\newpage

\section{Results of Country Rankings}
\label{sec:discussion}
\begin{figure}[H]
    \centering
    \includegraphics[width=1\linewidth]{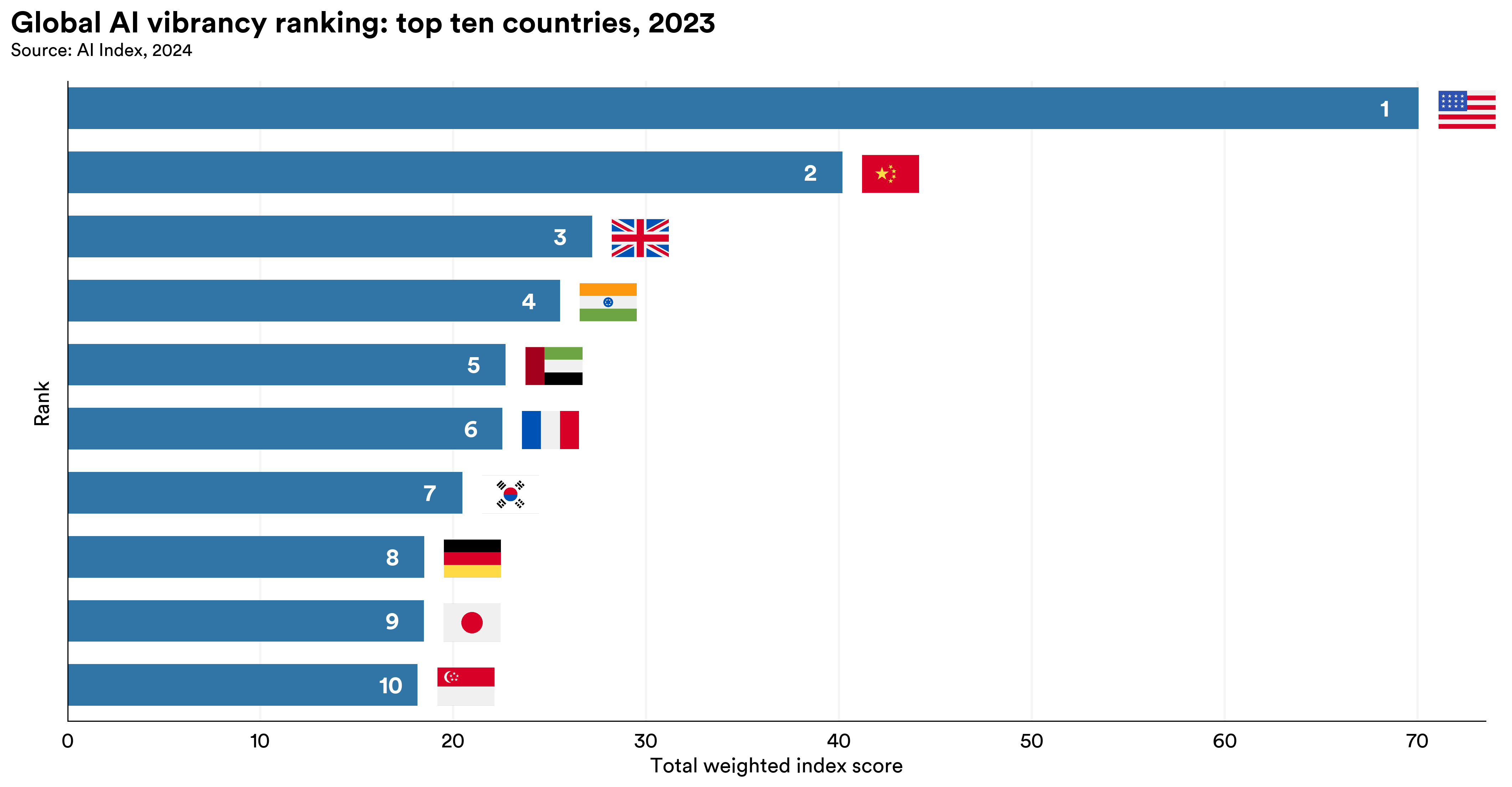}
    \caption{}
    \label{fig: vibrancy_ranking_top_10_countries_2023_total_score}
\end{figure}

\begin{figure}[H]
    \centering
    \includegraphics[width=1\linewidth]{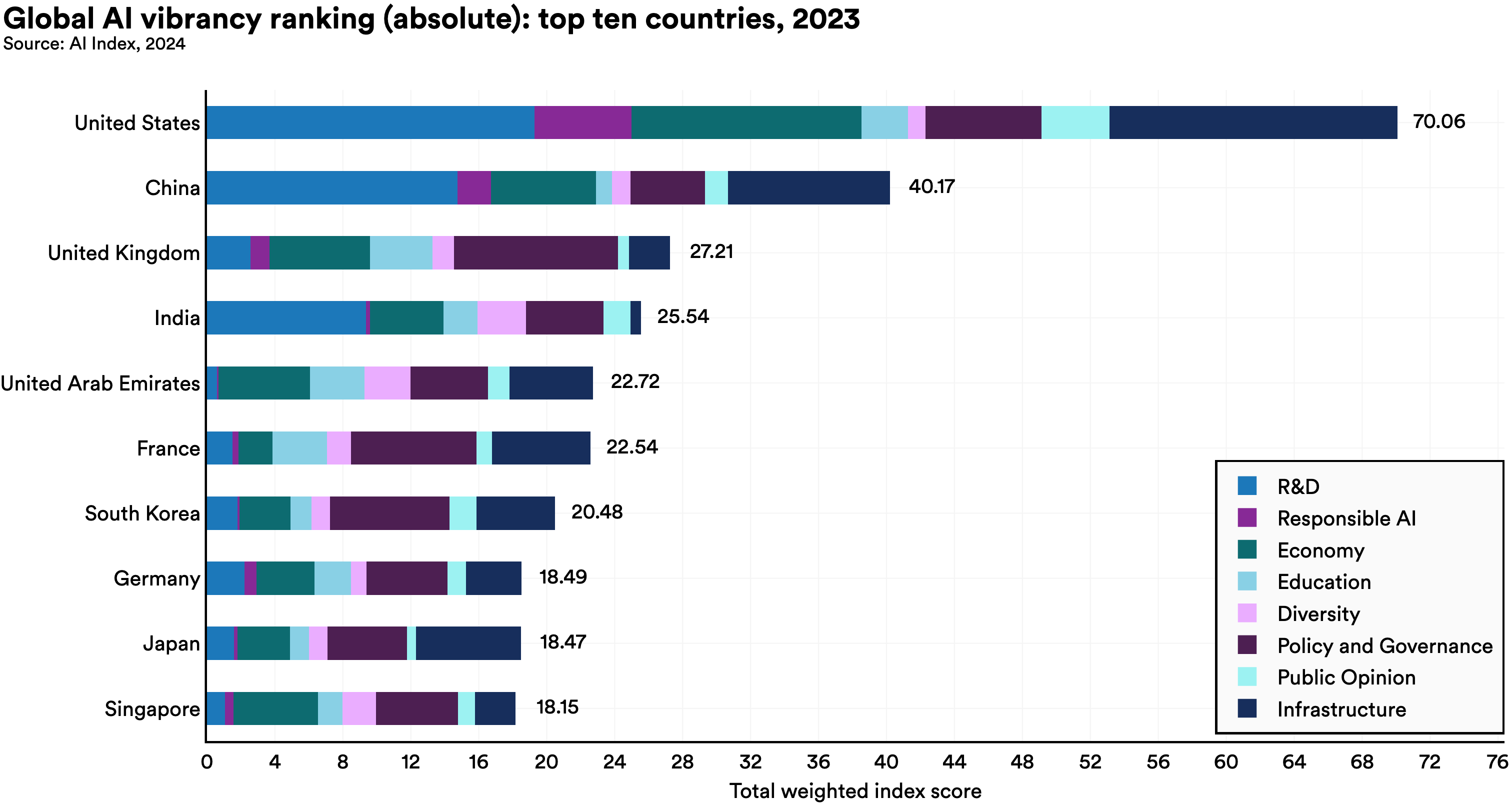}
    \caption{}
    \label{fig: global_ai_vibrancy_ranking_top10}
\end{figure}

In this section, we present the results of the 2023 Global AI Vibrancy Ranking. Figure \ref{fig: global_ai_vibrancy_ranking_top10} provides a snapshot of the current state of AI vibrancy in the top ten countries, evaluated across the eight key dimensions introduced in section \ref{sec:conceptual-framework}.\footnote{For a complete view of rankings using absolute values for all countries included in the analysis, refer to the Appendix \ref{sec:appendix-global_rankings_all_countries}, Figure \ref{fig: global_ai_vibrancy_ranking_all}.} Each country’s overall score is a weighted sum of these dimensions, computed using the absolute values of the metrics, reflecting their performance in AI-related activities and infrastructure.

The United States leads the 2023 ranking by a significant margin, with a total weighted index score of 70.06, reflecting its dominance across nearly all dimensions. The country’s strengths lie in its robust R\&D ecosystem, advanced infrastructure, and active policy and governance frameworks. The high scores in Economy and Responsible AI further solidify its position as a global leader in AI. The United States is the global leader in several key indicators of AI vibrancy. For instance, in 2023, it produced more notable machine learning models, had more total AI private investment, and submitted more responsible AI papers to leading conferences than any other country included in the analysis. Several notable AI companies, such as OpenAI, Meta, and Google, are headquartered in the United States. Additionally, many leading computer science universities, including Stanford, MIT, and Carnegie Mellon, are also based in the U.S. 

China holds the second position with a score of 40.17. It demonstrates substantial strengths in R\&D, Economy, and Infrastructure. China's focus on developing cutting-edge AI technologies and increasing its R\&D investments has positioned it as a major AI powerhouse. In 2023, China led the world in AI journal and conference publications. China's strength in research and development is unsurprising, given that many Chinese universities, such as the Chinese Academy of Sciences, Tsinghua University, and the University of Chinese Academy of Sciences, have traditionally produced a high number of AI-related publications \cite{maslej_artificial_2023}. Additionally, China is home to corporations like Baidu, which have developed significant large language models like Ernie 4.0. In terms of notable machine learning models and total private AI investment, China ranked second only to the United States. However, as detailed in the following section, despite China's strong position, there remains a significant gap between the two nations. The United States not only leads but also surpasses China (in absolute terms) by a considerable margin in many key metrics.

The United Kingdom ranks third with a score of 27.21, demonstrating particular strength in the Research and Development, Education, and Policy and Governance pillars. Home to top computer science universities like Oxford, Cambridge, and Imperial, the UK also hosts DeepMind, a Google subsidiary which in recent years has been a leader in AI research. Additionally, the UK leads in the number of AI study programs and, in 2023, had more mentions of AI in parliamentary proceedings than any other country. The UK has been quite politically engaged with AI, hosting the world's first international AI Safety Summit in 2023 \cite{uk_government_ai_2024} and was among the first countries to launch an institute dedicated to AI safety \cite{uk_government_introducing_2023}.

India's fourth-place position with 25.54 points is driven by strong performance in R\&D and recent improvements in the Economy pillar. India boasts a strong AI research community, ranking first in AI conference citations and third in total AI journal publications. Additionally, India ranked second globally in the number of AI GitHub projects. Public discourse around AI is also robust in India, as the country ranks second in both AI-related social media voice share and total AI social media posts.

The United Arab Emirates ranks fifth with a score of 22.72. The UAE ranked highly for the Economy pillar, as in 2023 it was among the top three nations globally for AI minority stake and public offering investments. It also scores highly on other economic indicators such as net migration of AI talent and has a fairly diverse AI workforce in terms of gender. In the last year, the UAE has also produced a significant number of notable machine learning as well as foundation models. On the infrastructure side, it reports very fast internet speed. The UAE's strong position in the vibrancy ranking highlights the Middle East's growing influence in the global AI landscape and is reflective of the UAE government's deliberate efforts to position the country as a significant global AI player \cite{perrigo_uae_2024}.

Other European countries, including France and Germany, feature prominently in the top ten, indicating Europe's collective commitment to AI development. Europe was among the first regions to pass significant AI legislation, with the EU enacting the AI Act in 2024. France ranks $6^{th}$ with 22.54 points, showing strength in Policy and Governance, Education, and Infrastructure. In 2023, France ranked second (tied with South Korea and the United Kingdom) in passing AI-related legislation. French policymakers have increasingly engaged in AI discussions, and the country has committed to hosting the next AI Safety Summit in 2025 \cite{pouget_frances_2024}. Additionally, Mistral, one of Europe's most prominent LLM developers, is headquartered in France and has contributed several notable machine learning models. Germany also fares well-coming in $8^{th}$: Germany is a significant contributor to AI research, ranking fourth in producing notable machine learning models. It has also published extensively on responsible AI and ranks fourth in total AI private investment.

Smaller countries like Singapore have made it into the top ten, suggesting that population and geographic scale are not the sole determinants of AI vibrancy.
The presence of South Korea ($7^{th}$ with 20.48 points) alongside Japan ($9^{th}$ with 18.47 points) and Singapore ($10^{th}$ with 18.15 points) in the top ten highlights the growing importance of AI in Asian economies. South Korea hosted the 2nd AI safety summit in 2024, and is home to many leading AI universities and companies, such as the Korea Advanced Institute of Science and Technology (KAIST) and NAVER.

The varied geographic representation in this ranking highlights the global nature of AI, as well as the diverse strategies different countries are adopting to stimulate growth and deployment in this field. Many national leaders have become aware of the geopolitical significance of AI and are working to improve levels of AI development within their own countries.

\begin{figure}[H]
    \centering
    \includegraphics[width=1\linewidth]{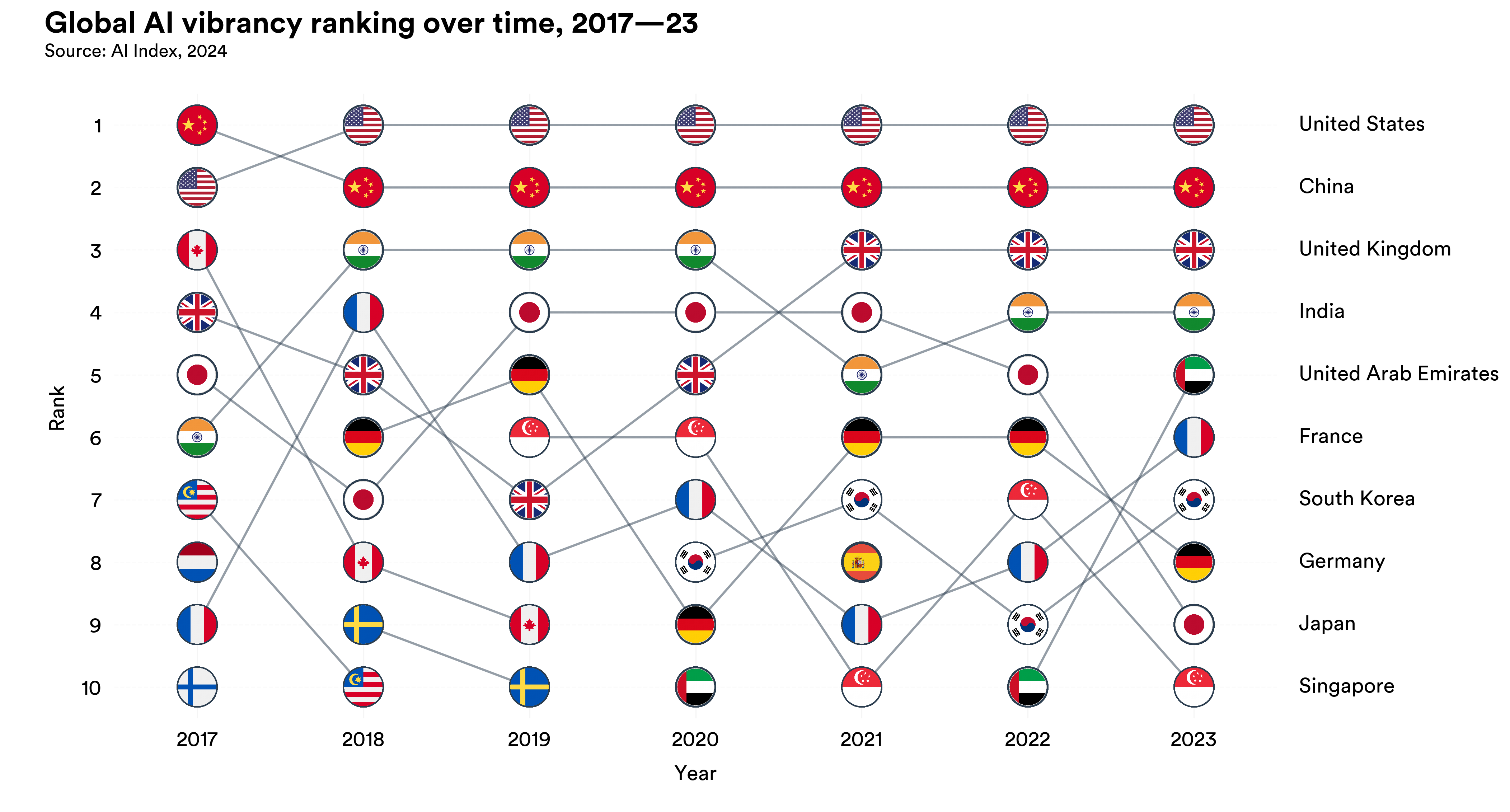}
    \caption{}
    \label{fig: vibrancy_ranking_top_10_countries_all_years_over_time_slope}
\end{figure}

Figure \ref{fig: vibrancy_ranking_top_10_countries_all_years_over_time_slope} illustrates the evolving competitive dynamics in global AI vibrancy between 2017 and 2023. It reveals three distinct tiers of competition: the top tier, consistently dominated by the United States and China; a more stable middle tier including the United Kingdom and India; and a more volatile lower tier where countries like France, Germany, Japan, Singapore, and South Korea frequently exchange positions.\footnote{Note that the interpretation of the rankings over time may be influenced by the availability of data for specific indicators. In
some cases, certain indicators may not be available for all years or countries, potentially affecting the comparability of
the index across time periods. For detailed information on data coverage, see the Appendix \ref{sec:appendix-coverage}.}

\begin{figure}[H]
    \centering
    \includegraphics[width=1\linewidth]{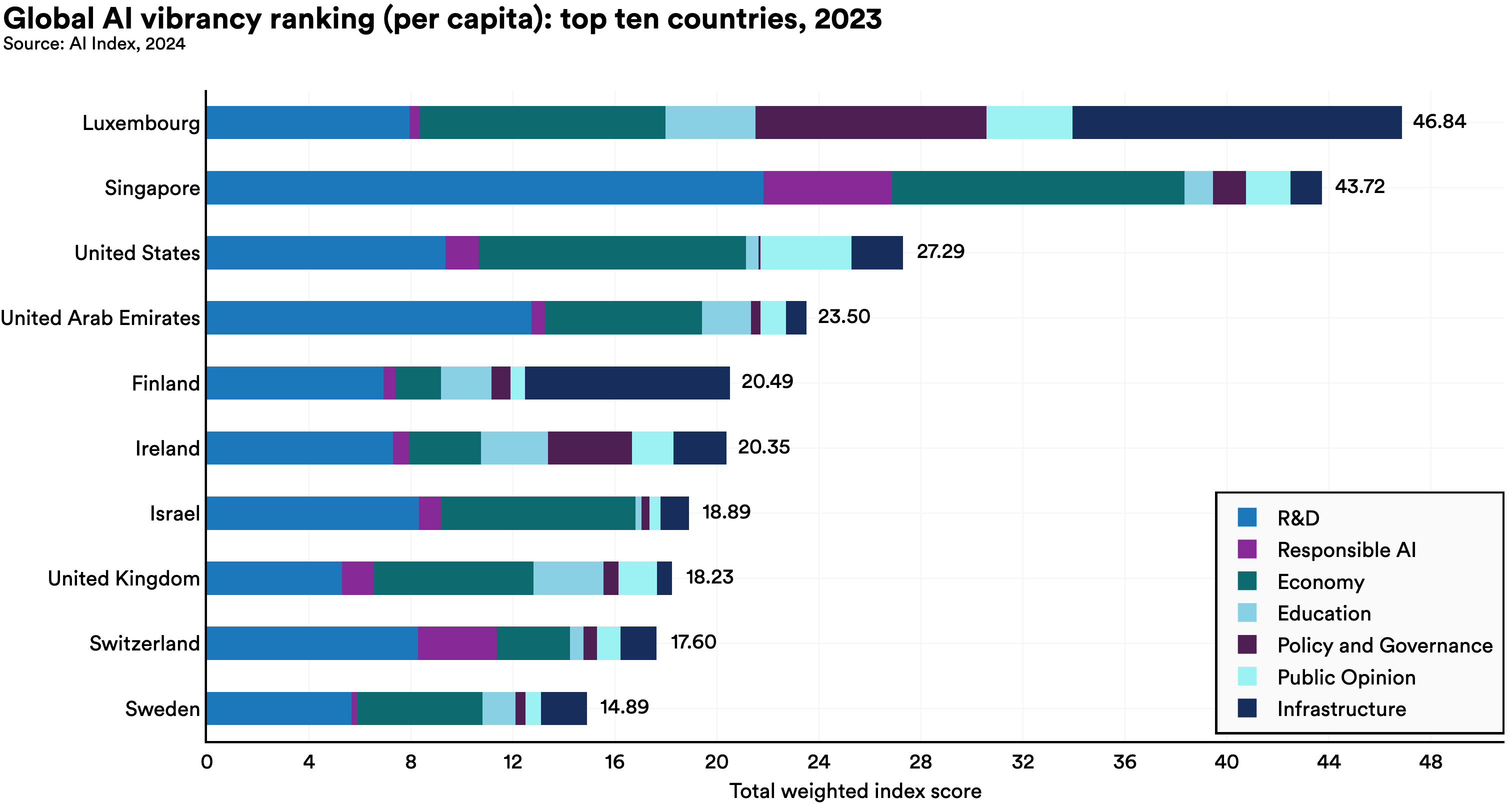}
    \caption{}
    \label{fig: global_ai_vibrancy_ranking_top10_percapita}
\end{figure}

The GVT also includes a ``per capita'' view that adjusts levels of AI vibrancy based on a country's population. When viewed from this perspective, Luxembourg holds the top position, achieving a total index score of 46.84. Luxembourg's strong per capita ranking is driven by balanced contributions across almost all dimensions, particularly Economy, Policy and Governance, and Infrastructure (Figure \ref{fig: global_ai_vibrancy_ranking_top10_percapita}). Singapore follows in second place with a score of 43.72, followed by the United States with a score of 27.29. Other countries that score especially well according to the per capita view are Finland, Ireland, Israel, the UK, Switzerland, and Sweden.\footnote{For a complete view of rankings using per capita values for all countries included in the analysis, refer to the Appendix \ref{sec:appendix-global_rankings_all_countries}, Figure \ref{fig: global_ai_vibrancy_ranking_all_percapita}.}

\subsection{Insights from the Sub-Indices}

\begin{figure}[H]
    \centering
    \begin{minipage}{0.48\linewidth}  
        \centering
        \includegraphics[height=5.5cm]{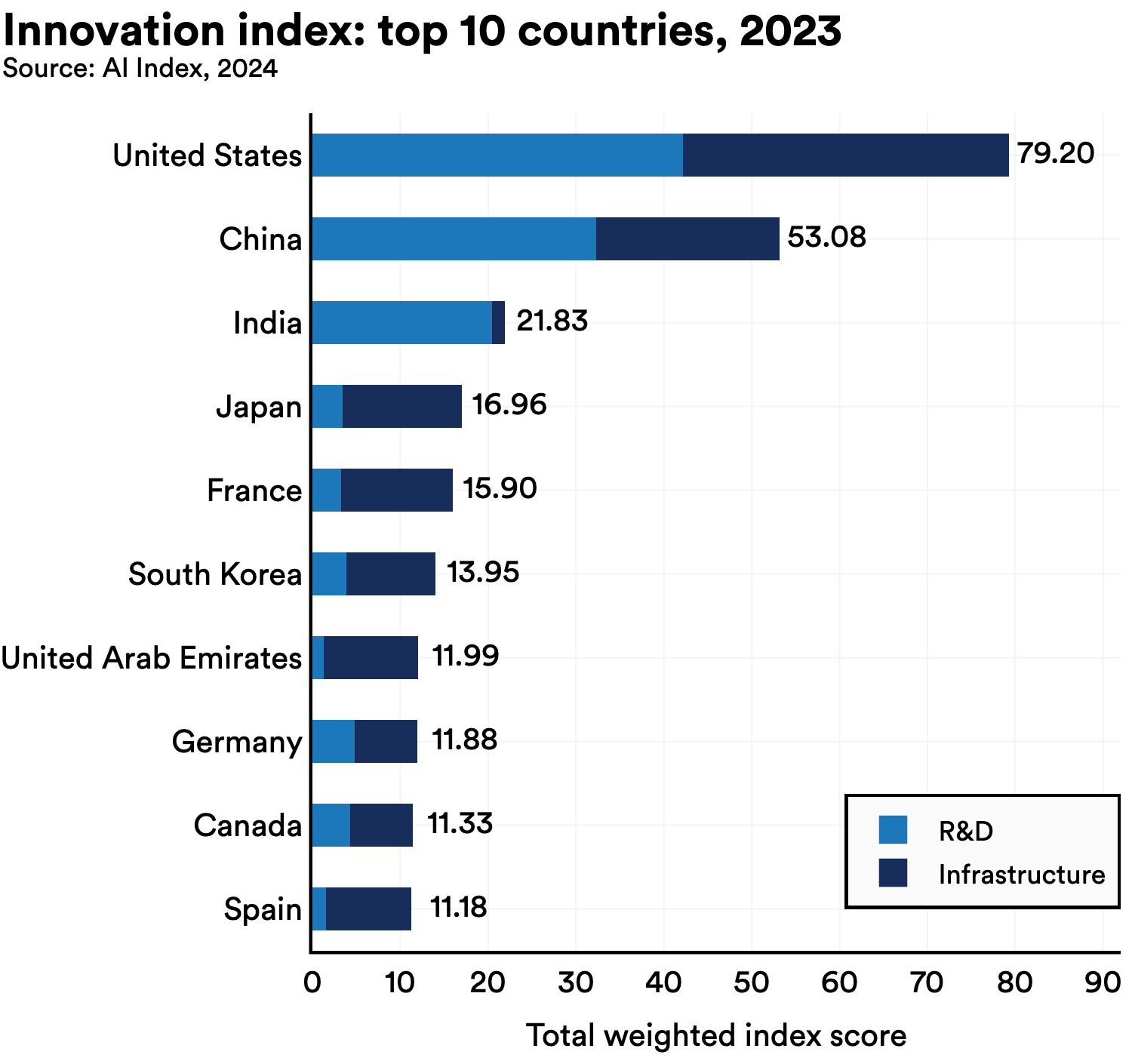}
        \caption{}
        \label{fig: innovation_ranking_top_10_countries_2023}
    \end{minipage}
    \hspace{0.02\linewidth}
    \begin{minipage}{0.48\linewidth}
        \centering
        \includegraphics[height=5.5cm]{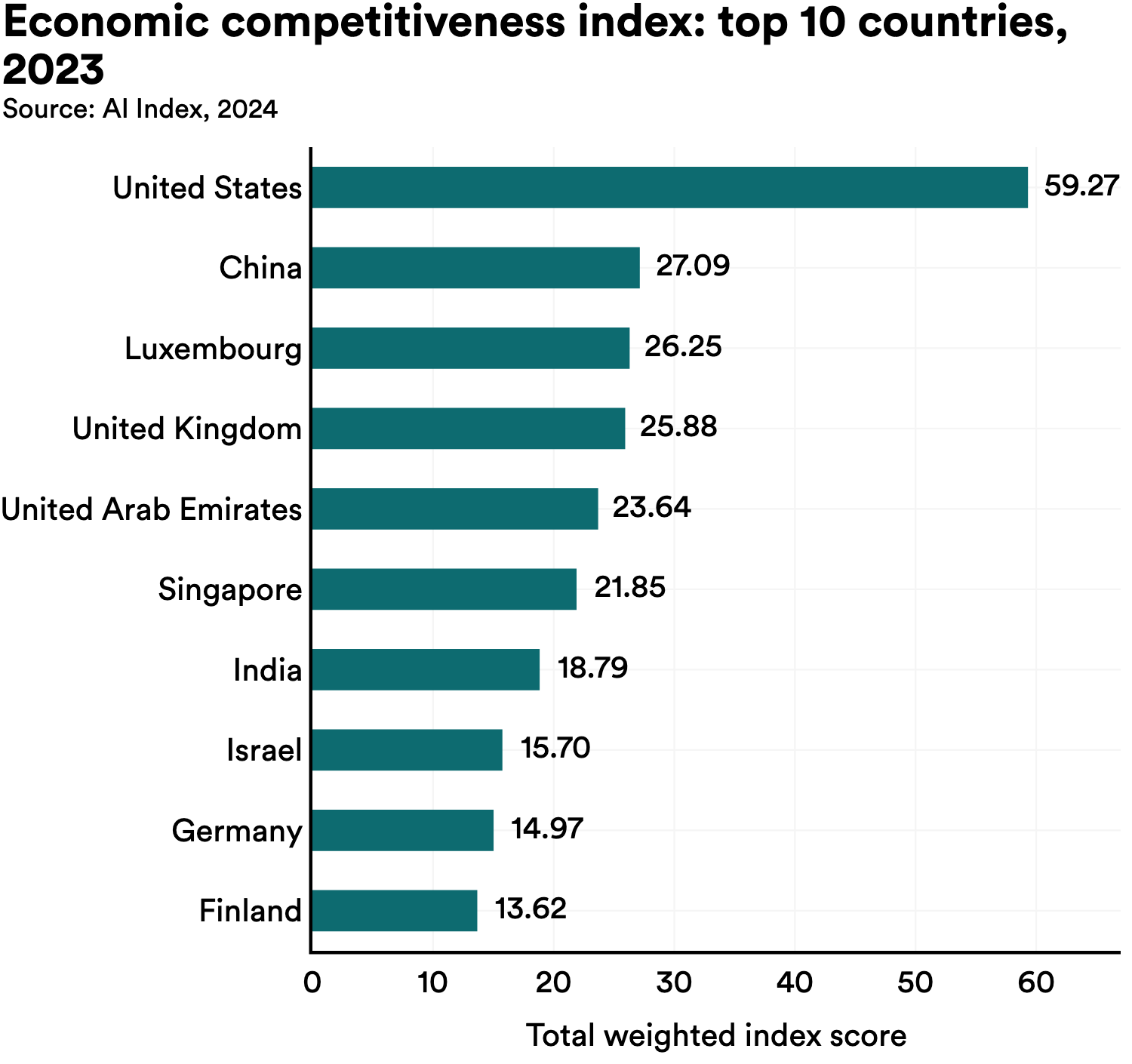}
        \caption{}
        \label{fig: economy_ranking_top_10_countries_2023}
    \end{minipage}
    
    \vspace{0.5cm}
    \begin{minipage}{0.48\linewidth}
        \centering
        \includegraphics[height=5.5cm]{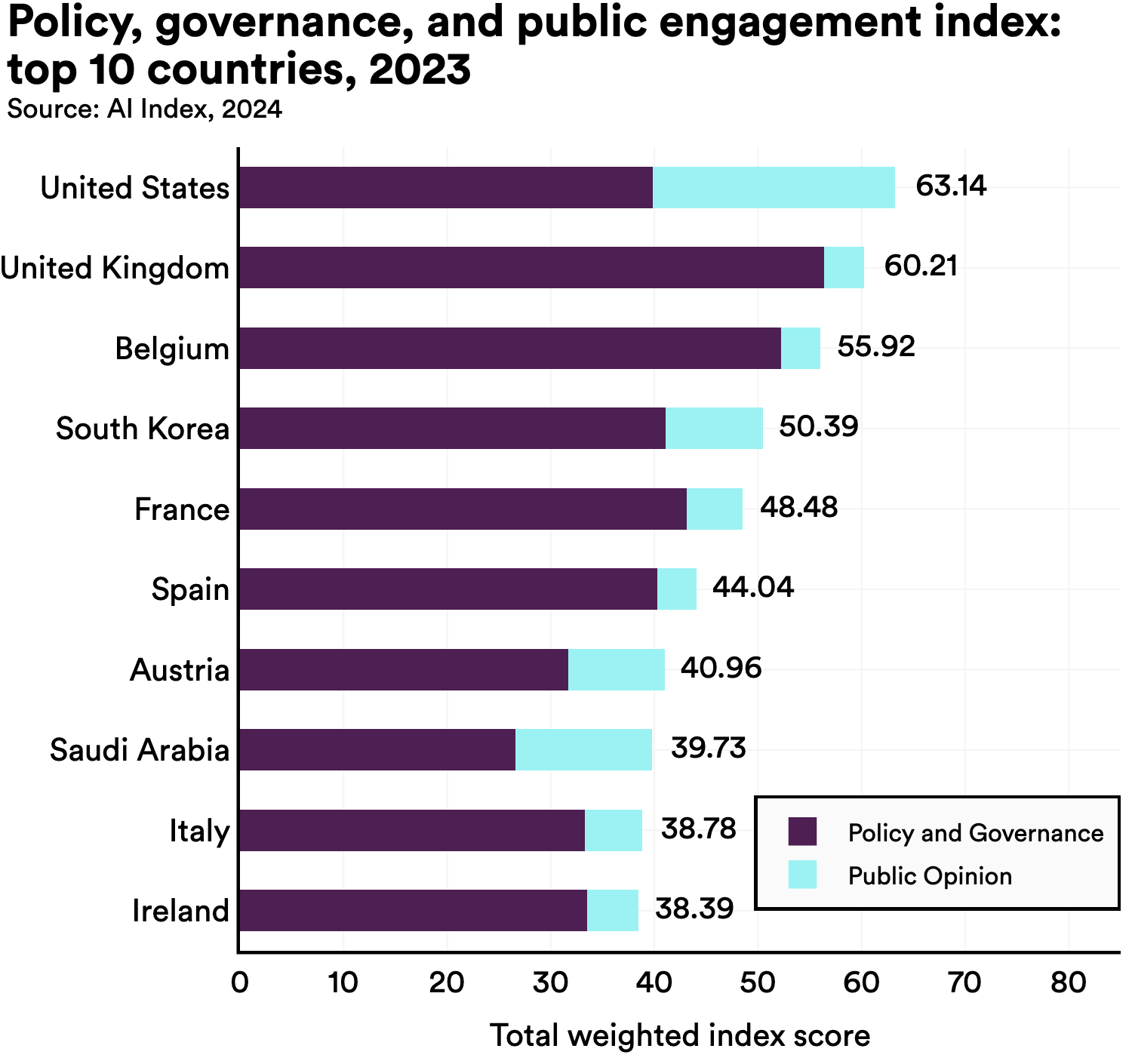}
        \caption{}
        \label{fig: policy_governance_public_opinion_ranking_top_10_countries_2023}
    \end{minipage}

\end{figure}

A closer examination of the AI Vibrancy Index's sub-indices introduced in Section \ref{subsec:sub-indices}, reveals a nuanced landscape where national strengths vary across dimensions. In the Innovation domain for 2023 (Figure \ref{fig: innovation_ranking_top_10_countries_2023}), the United States leads with a score of 79.20, substantially outperforming its nearest competitor, China (53.08). The U.S. especially excels in both R\&D and infrastructure. India comes in third place. The gap between Japan (16.96) the fourth place country and Spain, which comes in tenth (11.18) is fairly small in comparison to the gap that exists between the U.S. and China or U.S. and India. This contrast highlights a closer grouping among countries outside the top three, suggesting that modest investments in AI capacity could lead to substantial shifts in rankings.

The Economic Competitiveness sub-index for 2023 (Figure \ref{fig: economy_ranking_top_10_countries_2023}) captures a country's capacity to foster AI market dynamism, attract investments, and cultivate specialized talents. The United States again leads decisively with a score of 59.27. In the AI space, the U.S. leads in being able to attract AI investments, concentrate specialized talent, and generate employment opportunities. Luxembourg's third place position (26.25), coming narrowly behind China in second (27.09), is a remarkable achievement for a smaller nation. Luxembourg's relatively high ranking reflects the fact that it scores highly on several rate based metrics of AI economic vibrancy, such as AI hiring rate, AI talent concentration and the net migration of AI skills. For all of those indicators, Luxembourg occupies a top three global position. Singapore (21.85) and Israel (15.70) also perform strongly.

The Policy, Governance, and Public Engagement sub-index for 2023 (Figure \ref{fig: policy_governance_public_opinion_ranking_top_10_countries_2023}), which measures the intensity of AI-related policy activities, legislative initiatives, and public discourse, reveals the most geographically diverse leadership landscape among the three sub-indices. The United States again leads, with a score of 63.14, though its margin of advantage is narrower compared to other sub-indices. Close are the United Kingdom (60.21) and Belgium (55.92), which demonstrate similarly high levels of policy dynamism and stakeholder engagement. This tight clustering reflects the intense focus on AI governance across multiple regions, particularly in Europe, where legislative activity has increased in response to the rapid pace of AI development and deployment. South Korea (50.39) also scores well on this sub-index. Saudi Arabia’s emergence in the top ten, with a score of 39.73, signals growing momentum in AI public engagement and governance in the Middle East. The rise of countries in the Middle East further highlights how the scope of AI leadership is moving beyond traditional centers.\footnote{For a complete view of results for all countries included in the analysis, refer to the Appendix \ref{sec:appendix-sub_indices_rankings_all_countries}.}

\subsection{United States and China}

As AI has gained geopolitical importance, commentators and policymakers have increasingly focused on the comparative strengths of the two leading AI nations: the United States and China. This focus is evident in popular discussions about the geopolitical race between these countries and its implications, as well as in policy actions like the CHIPS and Science Act which were designed to bolster the U.S.-made semiconductor ecosystem \cite{tobin_china_2024, the_economist_america_2022, altchek_eric_2024}. Given this context, it is natural to closely compare the AI positioning of these two nations and examine how it has evolved over time.

\begin{figure}[H]
    \centering
    \begin{minipage}{0.48\linewidth}
        \centering
        \includegraphics[height=5.5cm]{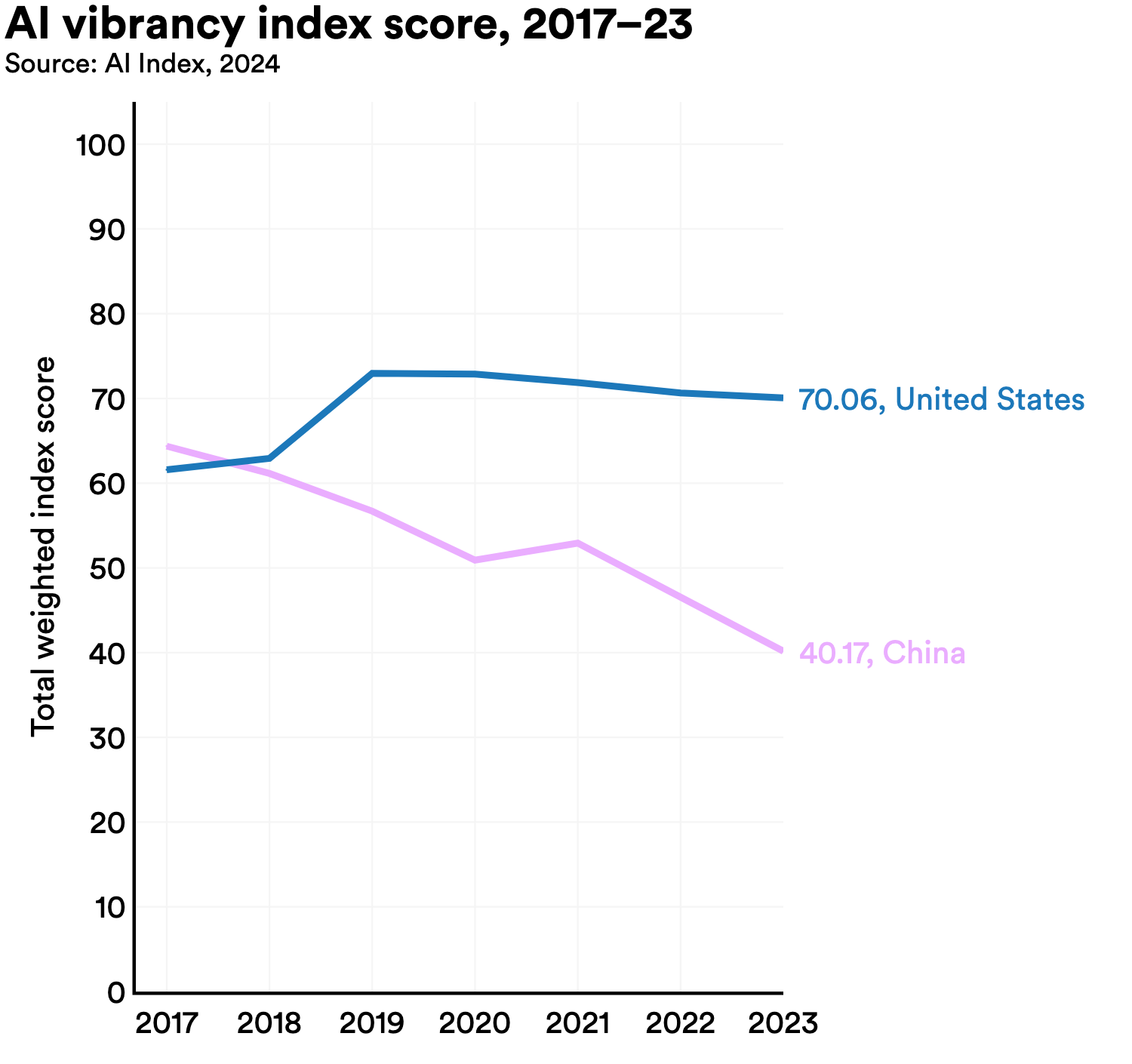}
        \caption{}
        \label{fig: vibrancy_score_us_china_over_time}
    \end{minipage}
    \hspace{0.02\linewidth}
    \begin{minipage}{0.48\linewidth}
        \centering
        \includegraphics[height=5.5cm]{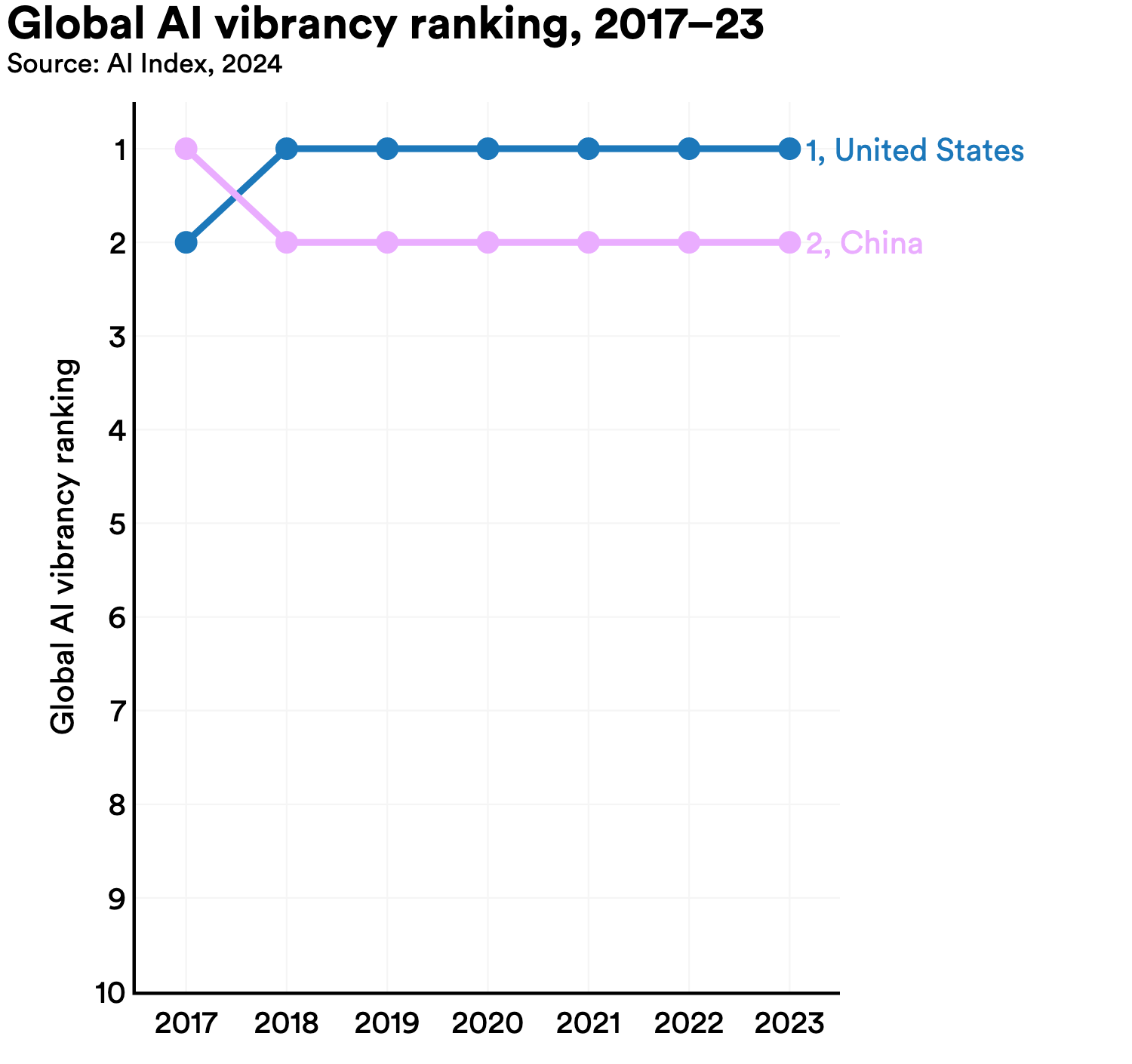}
        \caption{}
        \label{fig: vibrancy_ranking_us_china_over_time}
    \end{minipage}
    
    \vspace{0.5cm}
    \begin{minipage}{0.48\linewidth}
        \centering
        \includegraphics[height=5.5cm]{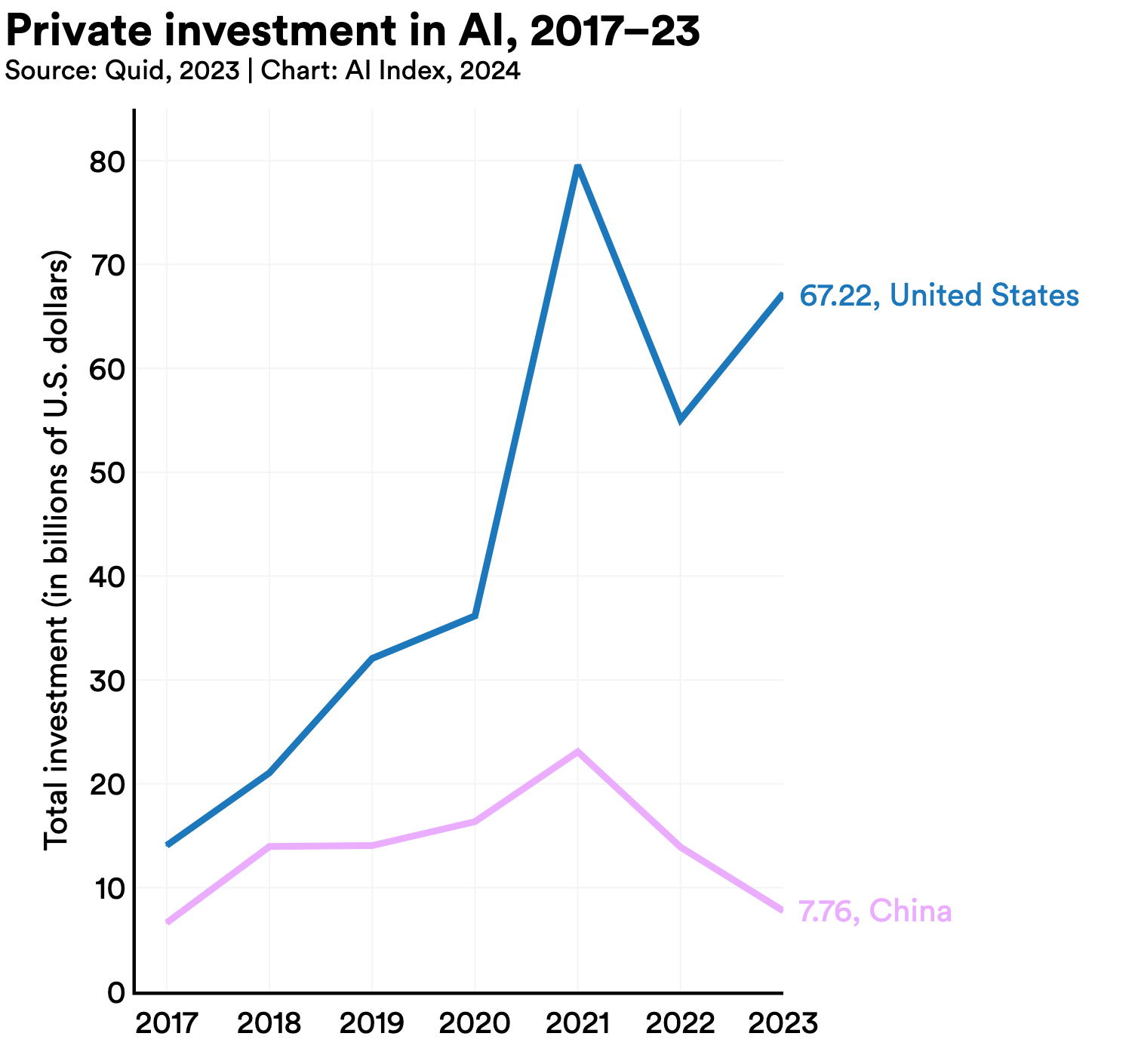}
        \caption{}
        \label{fig: ai_private_investments_us_vs_china_over_time}
    \end{minipage}
    \hspace{0.02\linewidth}
    \begin{minipage}{0.48\linewidth}
        \centering
        \includegraphics[height=5.5cm]{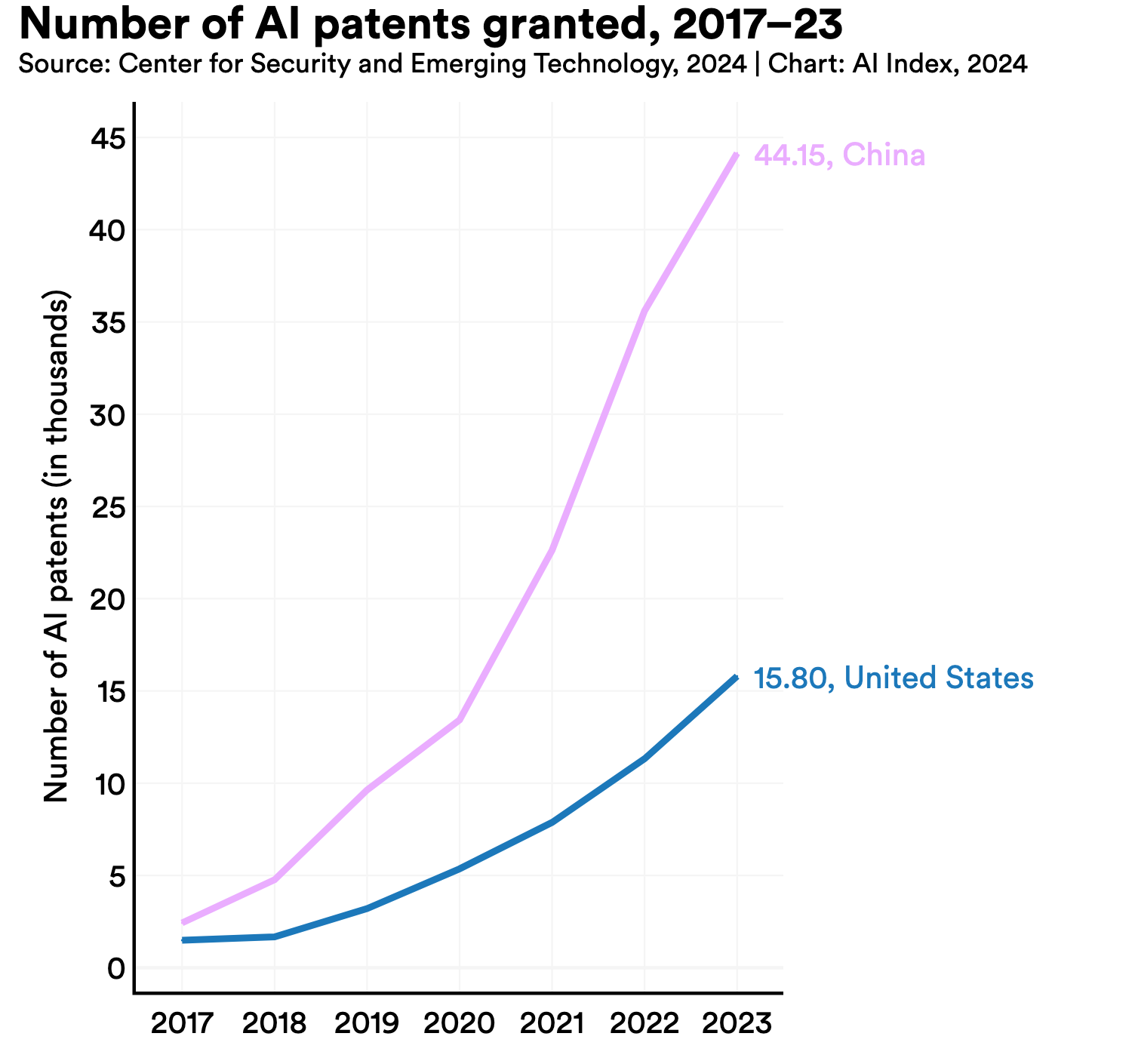}
        \caption{}
        \label{fig: num_patents_granted_us_vs_china_over_time}
    \end{minipage}
    
    \vspace{0.5cm}
     \begin{minipage}{0.48\linewidth}
        \centering
        \includegraphics[height=5.5cm]{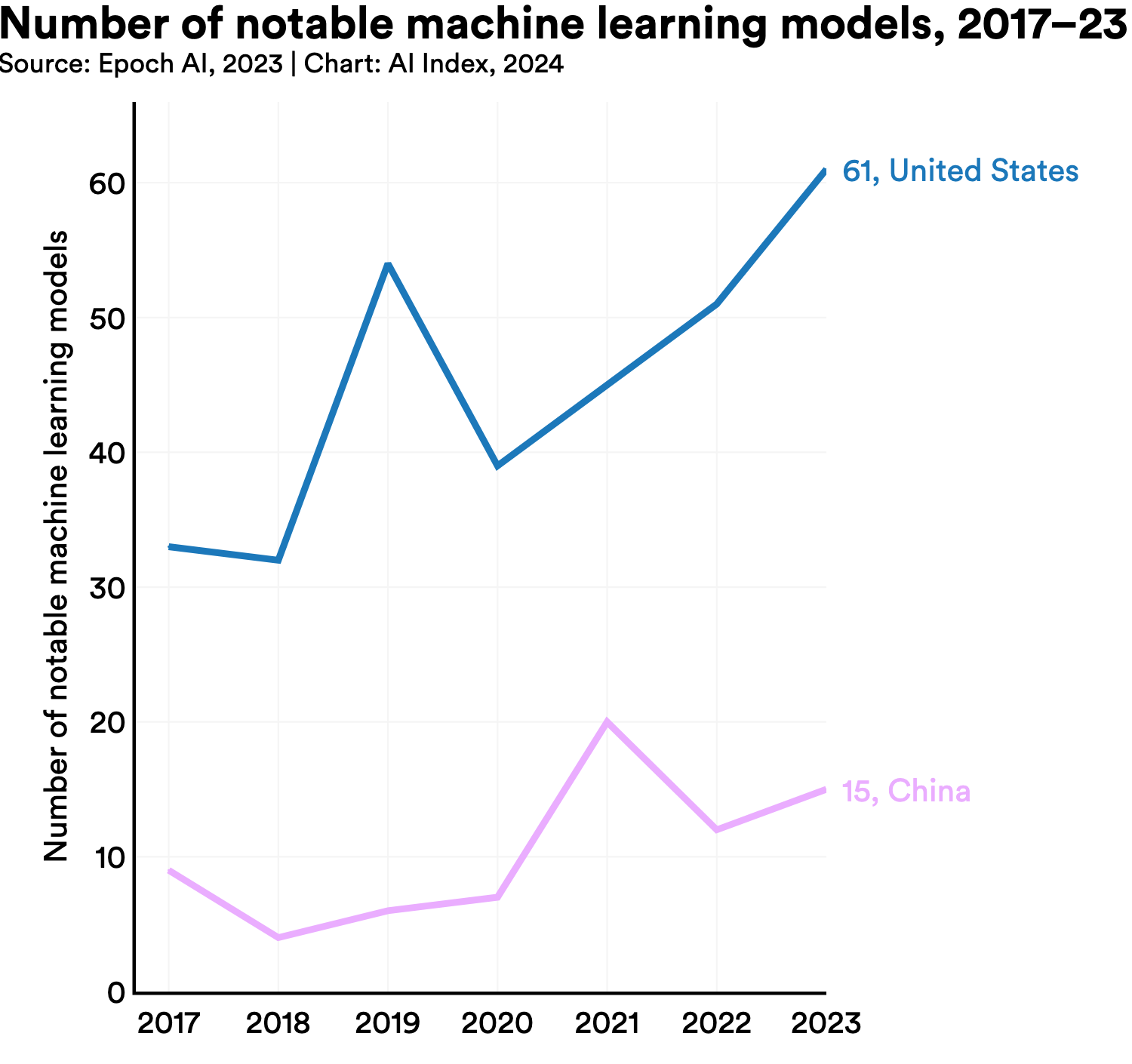}
        \caption{}
        \label{fig: num_notable_ml_models_over_time}
    \end{minipage}
    \begin{minipage}{0.48\linewidth}
        \centering
        \includegraphics[height=5.5cm]{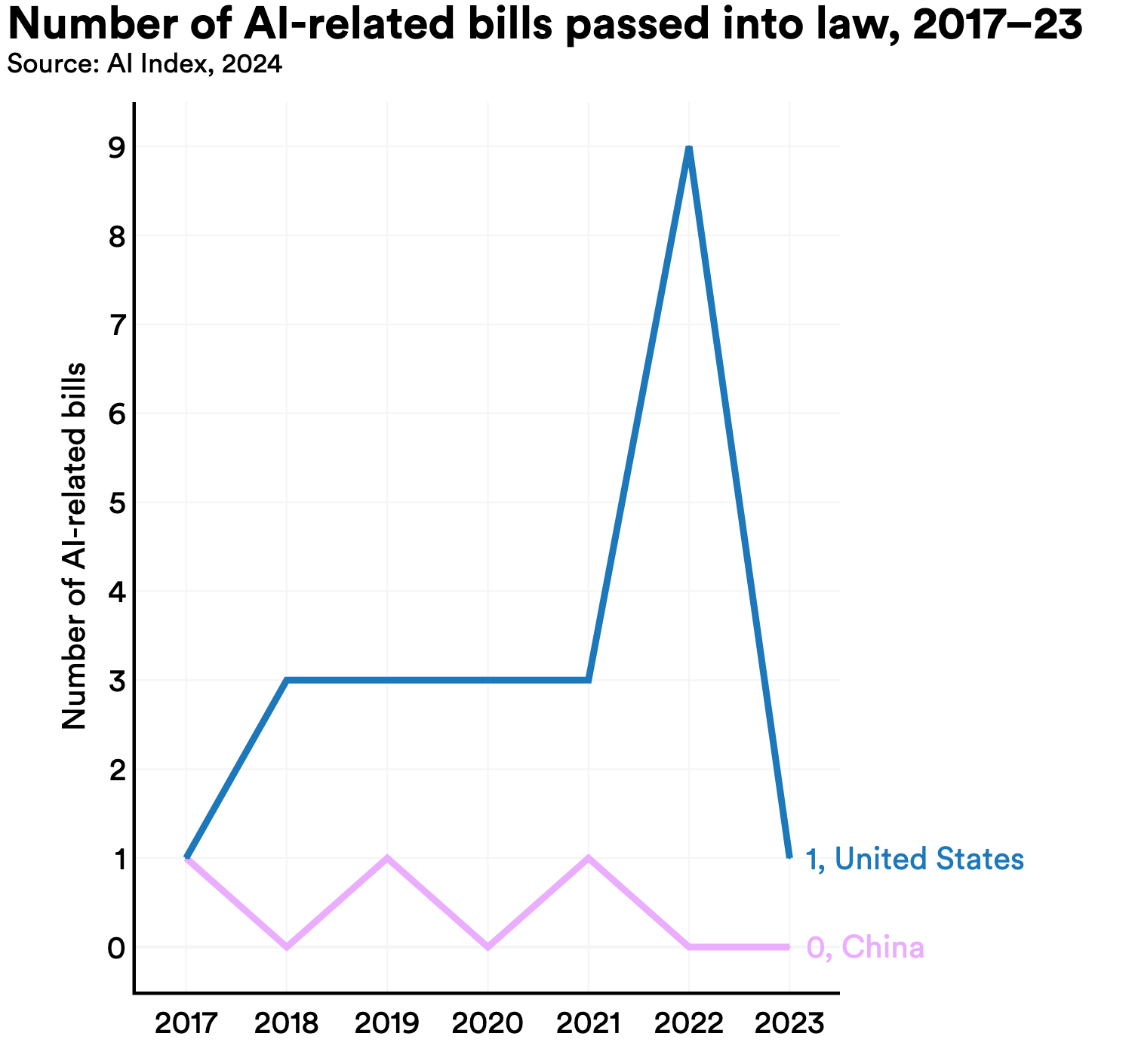}
        \caption{}
        \label{fig: num_ai_related_bills_passed_us_vs_china_over_time}
    \end{minipage}

\end{figure}

In 2018, the United States overtook China as the nation with the greatest global AI vibrancy. Since then, it has further reinforced its lead position.  In 2017, the two countries were relatively close in several key areas, including research and development and investment. However, by 2023, the United States had pulled ahead, achieving a vibrancy index score of 70.06, nearly double China's score of 40.17 (Figure \ref{fig: vibrancy_score_us_china_over_time}). 

The relative strength of the United States in comparison to China is evident when you look at a selection of significant AI indicators. The United States outpaced China in private investment, reaching \$67.22 billion in 2023 compared to China's \$7.76 billion (Figure \ref{fig: ai_private_investments_us_vs_china_over_time}). It also led in developing notable machine learning models, producing 61 models in 2023 compared to China's 15 (Figure \ref{fig: num_notable_ml_models_over_time}). However, China showed strong growth in AI innovation, particularly in patent generation, granting nearly three times as many AI patents as the United States (Figure \ref{fig: num_patents_granted_us_vs_china_over_time}). Additionally, the United States took a proactive stance in AI regulation, passing a total of 23 AI-related laws since 2017 (Figure \ref{fig: num_ai_related_bills_passed_us_vs_china_over_time}).\footnote{For more detailed methodological information on any specific indicator, refer to the most recent AI Index report \cite{maslej_ai_2024}.}

\subsection{Additional Notes and Limitations}

When interpreting the AI vibrancy rankings, it is essential to consider a few nuances. As outlined in the \nameref{sec:methodology} section, the rankings heavily rely on the weighting schema applied to each pillar and indicator. The AI Index has selected a weighting approach that it believes best represents the significance and coverage of the various components.\footnote{For more details on how weights were attributed, refer to the default weights selected for pillars and indicators in Appendix, Section \ref{sec:appendix-weighting}.} However, users of the vibrancy tool may have different views on the relative importance of specific pillars or indicators. The AI Index encourages users to explore the weight adjuster feature in the live tool to tailor the rankings to their own perspectives.

For many countries outside the top two or three in any given index in this paper, the score gaps between neighboring countries are relatively small. For instance, while the gap between the United States and China (the top two countries) was roughly 30 points on the weighted index score, the difference between the United Kingdom and Singapore (ranked third and tenth) was only about 9 points. As such, the rankings for countries outside the top positions (e.g., top three or ten) should be interpreted with more flexibility than those in the top few spots. Countries in the middle or lower tiers of the Index are often closely grouped, meaning that slight adjustments in weighting can significantly affect their ranks. This close clustering also indicates that countries can improve their standings through active policies and strategic planning that directly improve the strength of their AI ecosystem. 

Since the AI Index provides annual rankings, it is also important to recognize that a country's yearly position can be sensitive to outlier values. A high or low score in certain indicators one year might place a country out of alignment with its historical rankings. Future versions of the AI Vibrancy tool will include three- or five-year weighted averages to offer a more stable view of each country’s average position over time.

Finally, the AI Index recognizes that some data sources contributing to the rankings provide more comprehensive coverage for certain types of nations than others. To address this potential imbalance, the AI Index has down-weighted indicators or pillars with limited coverage. Efforts are underway to expand data coverage, and the AI Index encourages collaboration from governments and representatives worldwide to improve the representativeness of these rankings.

\section{Conclusion}
\label{sec:conclusion}
The Global AI Vibrancy Tool (GVT) is a robust and versatile platform for assessing and comparing AI vibrancy across countries.  By incorporating a comprehensive set of indicators across various dimensions, it provides a nuanced and dynamic understanding of over time trends in AI development at the national level. This tool is intended to serve as a valuable resource for policymakers, industry leaders, researchers, and the general public in guiding their understanding of the geopolitical dynamics that surround the AI ecosystem. The interactivity and customizability of the Global Vibrancy Tool allows users to explore and interpret AI data in meaningful ways, which can facilitate informed, flexible and strategic decision-making. As AI continues to evolve rapidly, the GVT will grow and expand in scope, with the intention of remaining an essential resource for tracking progress and identifying areas for improvement.

Future work will focus on expanding the range of indicators and countries, and improving data collection methods to ensure greater accuracy and coverage. Additionally, incorporating user feedback and enabling more sophisticated analytics capabilities will be important in maintaining the tool's relevance and utility.\footnote{Note that the AI Index may introduce new indicators and modify the interface throughout the year. Additional data for the new year will be incorporated during the summer following the release of the AI Index report. Any changes will be communicated on the AI Index website.} Collaborations with international organizations and AI research communities will also be pursued to enrich the tool's data sources and validation processes. These efforts are in the service of establishing the GVT as a comprehensive and authoritative measure of global AI vibrancy.

We encourage the AI community to suggest indicators and contribute data sources that are currently missing and require more effort to be collected. This collective input will be valuable in enhancing the tool's comprehensiveness and ensuring it meets the evolving needs of all stakeholders.

\newpage

\begin{acks}
We express our sincere gratitude to all who contributed to the development of the Global AI Vibrancy Tool (GVT). Our thanks go to Quid, LinkedIn, Lightcast, Studyportals, Epoch AI, Center for Security and Emerging Technology, and CRFM for providing the data. We are grateful to Andrew Shi for the analysis of responsible AI-related conference submissions and to Alexandra Rome, Julia Betts Lotufo, Njenga Kariuki and Naveen Venkatanarayanan for their valuable feedback. We thank Rishi Bommasani and Robi Rahman for insightful discussions. We thank the Steering Committee members—Erik Brynjolfsson, Jack Clark, Terah Lyons, James Manyika, Juan Carlos Niebles, and Russell Wald—for their ongoing guidance.  We thank Digital Avenues for their web development work, Travis Tyler for the web page design, and Shana Lynch and Jeanina Matias for their efforts in publicizing this project.

\vspace{12pt}

\noindent
The AI Index acknowledges that while authored by a team of human researchers, its writing process was aided by AI tools. Specifically, the authors used ChatGPT and Claude to help tighten and copy edit initial drafts.
\end{acks}

\bibliography{references}

\begin{thebibliography}{37}
\providecommand{\natexlab}[1]{#1}
\providecommand{\url}[1]{\texttt{#1}}
\expandafter\ifx\csname urlstyle\endcsname\relax
  \providecommand{\doi}[1]{doi: #1}\else
  \providecommand{\doi}{doi: \begingroup \urlstyle{rm}\Url}\fi

\bibitem[noa()]{noauthor_wdi_nodate}
{WDI} - {The} {World} by {Income} and {Region}.
\newblock URL \url{https://datatopics.worldbank.org/world-development-indicators/the-world-by-income-and-region.html}.

\bibitem[uni(1990)]{united_nations_human_1990}
Human {Development} {Report}.
\newblock Technical report, United Nations Development Programme (UNDP), New York Oxford University Press, 1990.
\newblock URL \url{https://hdr.undp.org/system/files/documents/hdr1990encompletenostats.pdf}.

\bibitem[noa(2024{\natexlab{a}})]{noauthor_live_2024}
Live data from {OECD}.{AI}, 2024{\natexlab{a}}.
\newblock URL \url{https://oecd.ai/en/data}.

\bibitem[noa(2024{\natexlab{b}})]{noauthor_ooklas_2024}
Ookla’s {Open} {Data} {Initiative}, 2024{\natexlab{b}}.
\newblock URL \url{https://www.ookla.com/ookla-for-good/open-data}.

\bibitem[noa(2024{\natexlab{c}})]{noauthor_top500_2024}
{TOP500}, 2024{\natexlab{c}}.
\newblock URL \url{https://www.top500.org/statistics/list/}.

\bibitem[Agrawal et~al.(2019)Agrawal, Gans, and Goldfarb]{agrawal_economics_2019}
Ajay Agrawal, Joshua Gans, and Avi Goldfarb, editors.
\newblock \emph{The {Economics} of {Artificial} {Intelligence}: {An} {Agenda}}.
\newblock National {Bureau} of {Economic} {Research} {Conference} {Report}. University of Chicago Press, Chicago, IL, May 2019.
\newblock ISBN 978-0-226-61333-8.
\newblock URL \url{https://press.uchicago.edu/ucp/books/book/chicago/E/bo35780726.html}.

\bibitem[AI(2024)]{epoch_ai_data_2024}
Epoch AI.
\newblock Data on {Notable} {AI} {Models}, 2024.
\newblock URL \url{https://epochai.org/data/notable-ai-models#explore-the-data}.

\bibitem[Altchek(2024)]{altchek_eric_2024}
Ana Altchek.
\newblock Eric {Schmidt} says {China} can't catch up to {US} in {AI} for 4 reasons.
\newblock \emph{Business Insider}, May 2024.
\newblock URL \url{https://www.businessinsider.com/eric-schmidt-comments-china-behind-united-states-ai-2024-5}.

\bibitem[Amodei and Hernandez(2018)]{amodei_ai_2018}
Dario Amodei and Danny Hernandez.
\newblock {AI} and {Compute}, 2018.
\newblock URL \url{https://openai.com/index/ai-and-compute/}.

\bibitem[Archibugi et~al.(2009)Archibugi, Denni, and Filippetti]{archibugi_technological_2009}
Daniele Archibugi, Mario Denni, and Andrea Filippetti.
\newblock The technological capabilities of nations: {The} state of the art of synthetic indicators.
\newblock \emph{Technological Forecasting and Social Change}, 76\penalty0 (7):\penalty0 917--931, September 2009.
\newblock URL \url{https://linkinghub.elsevier.com/retrieve/pii/S0040162509000043}.

\bibitem[Bommasani et~al.(2023)Bommasani, Soylu, Liao, Creel, and Liang]{bommasani_ecosystem_2023}
Rishi Bommasani, Dilara Soylu, Thomas~I. Liao, Kathleen~A. Creel, and Percy Liang.
\newblock Ecosystem {Graphs}: {The} {Social} {Footprint} of {Foundation} {Models}, March 2023.
\newblock URL \url{http://arxiv.org/abs/2303.15772}.

\bibitem[Cazzaniga et~al.(2024)Cazzaniga, Jaumotte, Li, Melina, Panton, Pizzinelli, Rockall, and Tavares]{cazzaniga_gen-ai_2024}
Mauro Cazzaniga, Florence Jaumotte, Longji Li, Giovanni Melina, Augustus~J. Panton, Carlo Pizzinelli, Emma Rockall, and Marina~M. Tavares.
\newblock Gen-{AI}: {Artificial} {Intelligence} and the {Future} of {Work}, 2024.
\newblock URL \url{https://www.imf.org/en/Publications/Staff-Discussion-Notes/Issues/2024/01/14/Gen-AI-Artificial-Intelligence-and-the-Future-of-Work-542379}.

\bibitem[Cesareo and White()]{cesareo_global_nodate}
Serena Cesareo and Joseph White.
\newblock The {Global} {AI} {Index}.
\newblock URL \url{https://www.tortoisemedia.com/intelligence/global-ai/}.
\newblock Tortoise Media.

\bibitem[Desai et~al.(2002)Desai, Fukuda-Parr, Johansson, and Sagasti]{desai_measuring_2002}
Meghnad Desai, Sakiko Fukuda-Parr, Claes Johansson, and Fransisco Sagasti.
\newblock Measuring the {Technology} {Achievement} of {Nations} and the {Capacity} to {Participate} in the {Network} {Age}.
\newblock \emph{Journal of Human Development}, 3\penalty0 (1):\penalty0 95--122, February 2002.
\newblock URL \url{https://doi.org/10.1080/14649880120105399}.

\bibitem[Djeffal et~al.(2022)Djeffal, Siewert, and Wurster]{djeffal_role_2022}
Christian Djeffal, Markus~B. Siewert, and Stefan Wurster.
\newblock Role of the state and responsibility in governing artificial intelligence: a comparative analysis of {AI} strategies.
\newblock \emph{Journal of European Public Policy}, 29\penalty0 (11):\penalty0 1799--1821, November 2022.
\newblock ISSN 1350-1763.
\newblock URL \url{https://doi.org/10.1080/13501763.2022.2094987}.

\bibitem[Economist(2022)]{the_economist_america_2022}
The Economist.
\newblock America takes on {China} with a giant microchips bill.
\newblock \emph{The Economist}, July 2022.
\newblock URL \url{https://www.economist.com/united-states/2022/07/29/america-takes-on-china-with-a-giant-microchips-bill}.

\bibitem[Furman et~al.(2002)Furman, Porter, and Stern]{furman_determinants_2002}
Jeffrey~L. Furman, Michael~E. Porter, and Scott Stern.
\newblock The determinants of national innovative capacity.
\newblock \emph{Research Policy}, 31\penalty0 (6):\penalty0 899--933, 2002.
\newblock ISSN 0048-7333.
\newblock URL \url{https://www.sciencedirect.com/science/article/pii/S0048733301001524}.

\bibitem[Gaulier and Zignago(2010)]{gaulier_baci_2010}
Guillaume Gaulier and Soledad Zignago.
\newblock {BACI}: {International} {Trade} {Database} at the {Product}-{Level} (the 1994-2007 {Version}).
\newblock \emph{SSRN Electronic Journal}, 2010.
\newblock ISSN 1556-5068.
\newblock URL \url{http://www.ssrn.com/abstract=1994500}.

\bibitem[Government(2023)]{uk_government_introducing_2023}
UK~Government.
\newblock Introducing the {AI} {Safety} {Institute}.
\newblock \emph{UK Department for Science, Innovation and Technology}, 2023.

\bibitem[Government(2024)]{uk_government_ai_2024}
UK~Government.
\newblock {AI} {Safety} {Summit} 2023 - {GOV}.{UK}, February 2024.
\newblock URL \url{https://www.gov.uk/government/topical-events/ai-safety-summit-2023}.

\bibitem[Greco et~al.(2019)Greco, Ishizaka, Tasiou, and Torrisi]{greco_methodological_2019}
Salvatore Greco, Alessio Ishizaka, Menelaos Tasiou, and Gianpiero Torrisi.
\newblock On the {Methodological} {Framework} of {Composite} {Indices}: {A} {Review} of the {Issues} of {Weighting}, {Aggregation}, and {Robustness}.
\newblock \emph{Social Indicators Research}, 141\penalty0 (1):\penalty0 61--94, January 2019.
\newblock URL \url{https://doi.org/10.1007/s11205-017-1832-9}.

\bibitem[Hankins et~al.()Hankins, Fuentes~Nettel, Martinescu, Grau, and Rahim]{hankins_2023_nodate}
Emma Hankins, Pablo Fuentes~Nettel, Livia Martinescu, Gonzalo Grau, and Sulamaan Rahim.
\newblock 2023 {Government} {AI} {Readiness} {Index}.
\newblock URL \url{https://oxfordinsights.com/ai-readiness/ai-readiness-index/}.

\bibitem[Incekara et~al.(2017)Incekara, Guz, and Sengun]{incekara_measuring_2017}
Ahmet Incekara, Tugba Guz, and Gulden Sengun.
\newblock Measuring the technology achievement index: comparison and ranking of countries.
\newblock \emph{Journal of Economics, Finance and Accounting}, 4\penalty0 (2):\penalty0 164--174, June 2017.
\newblock URL \url{http://www.pressacademia.org/images/documents/jefa/archives/vol_4_issue_2/11.pdf}.

\bibitem[Maslej et~al.(2023)Maslej, Fattorini, Brynjolfsson, Etchemendy, Ligett, Lyons, Manyika, Ngo, Niebles, Parli, Shoham, Wald, Clark, and Perrault]{maslej_artificial_2023}
Nestor Maslej, Loredana Fattorini, Erik Brynjolfsson, John Etchemendy, Katrina Ligett, Terah Lyons, James Manyika, Helen Ngo, Juan~Carlos Niebles, Vanessa Parli, Yoav Shoham, Russell Wald, Jack Clark, and Raymond Perrault.
\newblock Artificial {Intelligence} {Index} {Report} 2023, October 2023.
\newblock URL \url{http://arxiv.org/abs/2310.03715}.
\newblock arXiv:2310.03715 [cs].

\bibitem[Maslej et~al.(2024)Maslej, Fattorini, Perrault, Parli, Reuel, Brynjolfsson, Etchemendy, Ligett, Lyons, Manyika, Niebles, Shoham, Wald, and Clark]{maslej_ai_2024}
Nestor Maslej, Loredana Fattorini, Raymond Perrault, Vanessa Parli, Anka Reuel, Erik Brynjolfsson, John Etchemendy, Katrina Ligett, Terah Lyons, James Manyika, Juan~Carlos Niebles, Yoav Shoham, Russell Wald, and Jack Clark.
\newblock The {AI} {Index} 2024 {Annual} {Report}, April 2024.
\newblock URL \url{https://arxiv.org/abs/2405.19522}.

\bibitem[Nardo et~al.(2005)Nardo, Saisana, Saltelli, and Tarantola]{nardo_tools_2005}
Michela Nardo, Michaela Saisana, Andrea Saltelli, and Stefano Tarantola.
\newblock Tools for {Composite} {Indicators} {Building}.
\newblock Technical report, European Comission, January 2005.
\newblock URL \url{https://publications.jrc.ec.europa.eu/repository/handle/JRC31473}.

\bibitem[OECD et~al.(2008)OECD, Union, and EC-JRC]{oecd_handbook_2008}
OECD, European Union, and EC-JRC.
\newblock \emph{Handbook on {Constructing} {Composite} {Indicators}: {Methodology} and {User} {Guide}}.
\newblock 2008.
\newblock URL \url{https://www.oecd-ilibrary.org/content/publication/9789264043466-en}.

\bibitem[Pedro et~al.(2019)Pedro, Subosa, Rivas, and Valverde]{pedro_artificial_2019}
Francesc Pedro, Miguel Subosa, Axel Rivas, and Paula Valverde.
\newblock Artificial {Intelligence} in {Education}: {Challenges} and {Opportunities} for {Sustainable} {Development}.
\newblock \emph{UNESCO}, 2019.
\newblock URL \url{https://repositorio.minedu.gob.pe/bitstream/handle/20.500.12799/6533/Artificial%20intelligence%20in%20education%20challenges%20and%20opportunities%20for%20sustainable%20development.pdf?sequence=1&isAllowed=y}.

\bibitem[Perrigo(2024)]{perrigo_uae_2024}
Billy Perrigo.
\newblock The {UAE} {Is} on a {Mission} to {Become} an {AI} {Power}, March 2024.
\newblock URL \url{https://time.com/6958369/artificial-intelligence-united-arab-emirates/}.

\bibitem[Pouget(2024)]{pouget_frances_2024}
Hadrien Pouget.
\newblock France’s {AI} {Summit} {Is} a {Chance} to {Reshape} {Global} {Narratives} on {AI}, July 2024.
\newblock URL \url{https://carnegieendowment.org/posts/2024/07/france-ai-summit-reshape-global-narrative?lang=en}.

\bibitem[Radu(2021)]{radu_steering_2021}
Roxana Radu.
\newblock Steering the governance of artificial intelligence: national strategies in perspective.
\newblock \emph{Policy and Society}, 40\penalty0 (2):\penalty0 178--193, April 2021.
\newblock ISSN 1449-4035, 1839-3373.
\newblock URL \url{https://academic.oup.com/policyandsociety/article/40/2/178/6509308}.

\bibitem[Saisana et~al.(2019)Saisana, Vertesy, Neves, Benavente, Becker, Moura, and Dominguez-Torreiro]{saisana_coin_2019}
Michaela Saisana, Daniel Vertesy, Ana Neves, Daniela Benavente, William Becker, Carlos Moura, and Marcos Dominguez-Torreiro.
\newblock \emph{{COIN} tool – {User} guide}.
\newblock European Commission, Joint Research Centre, 2019.
\newblock URL \url{https://data.europa.eu/doi/10.2760/523877}.

\bibitem[Sartori and Bocca(2023)]{sartori_minding_2023}
Laura Sartori and Giulia Bocca.
\newblock Minding the gap(s): public perceptions of {AI} and socio-technical imaginaries.
\newblock \emph{AI \& SOCIETY}, 38\penalty0 (2):\penalty0 443--458, April 2023.
\newblock ISSN 1435-5655.
\newblock URL \url{https://doi.org/10.1007/s00146-022-01422-1}.

\bibitem[Shoham(2017)]{shoham_toward_2017}
Yoav Shoham.
\newblock Toward the {AI} {Index}.
\newblock \emph{AI Magazine}, 38\penalty0 (4):\penalty0 71--77, 2017.
\newblock URL \url{https://onlinelibrary.wiley.com/doi/abs/10.1609/aimag.v38i4.2761}.

\bibitem[Tobin and Metz(2024)]{tobin_china_2024}
Meaghan Tobin and Cade Metz.
\newblock China {Is} {Closing} the {A}.{I}. {Gap} {With} the {United} {States}.
\newblock \emph{The New York Times}, July 2024.
\newblock ISSN 0362-4331.
\newblock URL \url{https://www.nytimes.com/2024/07/25/technology/china-open-source-ai.html}.

\bibitem[{UNDP (United Nations Development Programme)}(2010)]{undp_united_nations_development_programme_hdr_2010}
{UNDP (United Nations Development Programme)}.
\newblock {HDR} 2010 - {The} {Real} {Wealth} of {Nations}: {Pathways} to {Human} {Development}.
\newblock Technical report, 2010.
\newblock URL \url{https://EconPapers.repec.org/RePEc:hdr:report:hdr2010}.

\bibitem[Zowghi and da~Rimini(2023)]{zowghi_diversity_2023}
Didar Zowghi and Francesca da~Rimini.
\newblock Diversity and {Inclusion} in {Artificial} {Intelligence}, May 2023.
\newblock URL \url{http://arxiv.org/abs/2305.12728}.

\end{thebibliography}

\newpage

\begin{appendices}
\section*{APPENDIX}

\section{List of Countries}
\label{sec:appendix-countries}

To ensure the quality and representativeness of the analysis, a data coverage threshold of 70\%, averaged over the last three years, was applied. This threshold was chosen to balance the need for comprehensive data with the goal of including a diverse set of countries. However, Estonia, Malaysia, Mexico, Russia, Saudi Arabia, and Turkey did not meet this threshold but have been included due to their strategic importance such as geopolitical influence, economic impact, or contributions to global trends. Therefore, these countries are included, but users should interpret their results with caution. For more details, see tables \ref{tab:data_coverage_countries} and \ref{tab:data_coverage_indicators}.

\begin{table}[h!]
\caption{List of Countries}
\centering
\small
\begin{tabular}{l}
\toprule
\textbf{Country} \\
\midrule
Australia \\
Austria \\
Belgium \\
Brazil \\
Canada \\
China \\
Denmark \\
Estonia \\
Finland \\
France \\
Germany \\
India \\
Ireland \\
Israel \\
Italy \\
Japan \\
Luxembourg \\
Malaysia \\
Mexico \\
Netherlands \\
New Zealand \\
Norway \\
Poland \\
Portugal \\
Russia \\
Saudi Arabia \\
Singapore \\
South Africa \\
South Korea \\
Spain \\
Sweden \\
Switzerland \\
Turkey \\
United Arab Emirates \\
United Kingdom \\
United States \\
\bottomrule
\end{tabular}
\label{tab:country_list}
\end{table}

\newpage

\section{List of Indicators and Descriptions}
\label{sec:appendix-indicators-descriptions}

\begin{center}
\small
\captionof{table}{List of Indicators and Descriptions}
\begin{longtable}{>{\raggedright\arraybackslash}p{6cm}>{\raggedright\arraybackslash}p{7.5cm}}
\toprule
\textbf{Indicator} & \textbf{Description} \\
\midrule
\endfirsthead

\toprule
\textbf{Indicator} & \textbf{Description} \\
\midrule
\endhead

\midrule
\endfoot

\bottomrule
\endlastfoot

AI Journal Publications & Number of published AI journal publications in a given country. For more details, refer to the Appendix of the 2024 AI Index Report \cite{maslej_ai_2024}{}. \\
\hline
AI Conference Publications & Number of published AI conference publications in a given country. For more details, refer to the Appendix of the 2024 AI Index Report \cite{maslej_ai_2024}{}. \\
\hline
AI Journal Citations & Number of published AI journal citations in a given country. For more details, refer to the Appendix of the 2024 AI Index Report \cite{maslej_ai_2024}{}. \\
\hline
AI Conference Citations & Number of published AI conference citations in a given country. For more details, refer to the Appendix of the 2024 AI Index Report \cite{maslej_ai_2024}{}. \\
\hline
AI Patent Grants & Number of AI patent grants in a given country. For more details, refer to the Appendix of the 2024 AI Index Report \cite{maslej_ai_2024}{}. \\
\hline
Notable Machine Learning Models & Number of notable machine learning models in a given country. For more details, refer to Epoch AI's webpage \cite{epoch_ai_data_2024}{}. \\
\hline
Academia-Industry Model Production Concentration & Indicator quantifying the balance between the number of notable machine learning models produced by academia and industry within each country. For more details, refer to the Appendix, Section \ref{sec:appendix-metrics}. \\
\hline
Foundation Models & Number of foundation models in a given country. For more details, refer to the Appendix of the 2024 AI Index Report \cite{maslej_ai_2024}{}. \\
\hline
Foundation Models Datasets & Number of foundation models' datasets in a given country. For more details, refer to the Appendix of the 2024 AI Index Report \cite{maslej_ai_2024}{}. \\
\hline
Foundation Models Applications & Number of foundation models' applications in a given country. For more details, refer to the Appendix of the 2024 AI Index Report \cite{maslej_ai_2024}{}. \\
\hline
Open Access Foundation Models & The proportion of foundation models that are openly accessible compared to the total number of foundation models in a given country. \\
\hline
AI GitHub Projects & Total number of AI-related projects on GitHub in a given country, consisting of a collection of files such as source code, documentation, configuration files, and images that together form a software project. For more details, refer to the Appendix of the 2024 AI Index Report \cite{maslej_ai_2024}{}. \\
\hline
AI GitHub Projects Stars & Total number of stars for AI-related projects on GitHub in a given country. Users can ``star'' a repository to show interest, similar to liking a post on social media, indicating support for the project. For more details, refer to the Appendix of the 2024 AI Index Report \cite{maslej_ai_2024}{}. \\
\hline
AAAI Conference Submissions on RAI Topics & Number of responsible AI-related academic submissions to the AAAI conference in a given country.  For more details, refer to the Appendix of the 2024 AI Index Report \cite{maslej_ai_2024}{}. \\
\hline
AIES Conference Submissions on RAI Topics & Number of responsible AI-related academic submissions to the AIES conference in a given country. For more details, refer to the Appendix of the 2024 AI Index Report \cite{maslej_ai_2024}{}. \\
\hline
FAccT Conference Submissions on RAI Topics & Number of responsible AI-related academic submissions to the FAccT conference in a given country. For more details, refer to the Appendix of the 2024 AI Index Report \cite{maslej_ai_2024}{}. \\
\hline
ICLR Conference Submissions on RAI Topics & Number of responsible AI-related academic submissions to the ICLR conference in a given country. For more details, refer to the Appendix of the 2024 AI Index Report \cite{maslej_ai_2024}{}. \\
\hline
ICML Conference Submissions on RAI Topics & Number of responsible AI-related academic submissions to the ICML conference in a given country. For more details, refer to the Appendix of the 2024 AI Index Report \cite{maslej_ai_2024}{}.\\
\hline
NeurIPS Conference Submissions on RAI Topics & Number of responsible AI-related academic submissions to the NeurIPS conference in a given country. For more details, refer to the Appendix of the 2024 AI Index Report \cite{maslej_ai_2024}{}.\\
\hline
Total AI Private Investment & Total amount of private investment received for AI startups (nominal USD) in a given country. For more details, refer to the Appendix of the 2024 AI Index Report \cite{maslej_ai_2024}{}. \\
\hline
Total AI Merger/Acquisition Investment & Total amount of merger/acquisition investment received for AI startups (nominal USD) in a given country. For more details, refer to the Appendix of the 2024 AI Index Report \cite{maslej_ai_2024}{}. \\
\hline
Total AI Minority Stake Investment & Total amount of minority stake investment received for AI startups (nominal USD) in a given country. For more details, refer to the Appendix of the 2024 AI Index Report \cite{maslej_ai_2024}{}. \\
\hline
Total AI Public Offering Investment & Total amount of public offering investment received for AI startups (nominal USD) in a given country. For more details, refer to the Appendix of the 2024 AI Index Report \cite{maslej_ai_2024}{}.\\
\hline
Newly Funded AI Companies & Total number of newly funded AI companies in the given country. For more details, refer to the Appendix of the 2024 AI Index Report \cite{maslej_ai_2024}{}.\\
\hline
Relative AI Skill Penetration & This indicator measures the intensity of AI skills within a country. It involves computing skill frequencies from LinkedIn members (2015-2023), re-weighting them using a TF-IDF model to identify the top 50 representative skills (the ``skill genome''), and calculating the proportion of AI skills among these top skills. This rate indicates the prevalence and intensity of AI skills used by LinkedIn members in their jobs. To allow for skills penetration comparisons across countries, skill genomes are calculated and compared to a benchmark (e.g., global average). A ratio is then constructed between a country’s AI skills penetration and the benchmark, controlling for occupations. A relative AI skills penetration of 1.5 indicates that AI skills are 1.5 times more frequent than in the benchmark for similar occupations. For more details, refer to the Appendix of the 2024 AI Index Report \cite{maslej_ai_2024}{}. \\
\hline
AI Hiring Rate YoY Ratio & This indicator measures the year-over-year change in the AI Hiring Rate relative to the Overall Hiring Rate in the same country. Each month, the AI Hiring Rate is calculated and divided by the Overall Hiring Rate. The year-over-year change of this ratio is then determined, followed by calculating a 12-month moving average. Interpretation: In 2023, India saw a 16.8\% year-over-year growth in the ratio of AI talent hiring relative to overall hiring. For more details, refer to the Appendix of the 2024 AI Index Report \cite{maslej_ai_2024}{}. \\
\hline
AI Talent Concentration & This indicator identifies AI talent as LinkedIn members who have explicitly added AI skills to their profile or work in AI-related occupations. AI Talent Concentration is calculated using the counts of AI talent relative to the total number of LinkedIn members in a given country. It is important to note that these metrics may be influenced by the level of LinkedIn coverage in different countries and should be interpreted with caution. For more details, refer to the Appendix of the 2024 AI Index Report \cite{maslej_ai_2024}{}. \\
\hline
AI Job Postings (\% of Total) & The percentage of job postings that require AI skills in a given country. Lightcast collects postings from over 51,000 online job sites to develop a comprehensive, real-time portrait of labor market demand. For more details, refer to the Appendix of the 2024 AI Index Report \cite{maslej_ai_2024}{}. \\
\hline
Net Migration Flow of AI Skills & AI talent migration per 10,000 LinkedIn members in a given country. This metric tracks AI talent with AI skills or jobs. For a specific country (country A), net talent flows are calculated by dividing the net AI talent migration by the Member Count, normalized per 10,000 members. This indicates the relative migration of AI talent to and from country A. For more details, refer to the Appendix of the 2024 AI Index Report \cite{maslej_ai_2024}{}. \\
\hline
AI Study Programs in English & The number of English-language AI-related study programs in a given country. A study program, or degree program, comprises a series of courses designed to enable students to earn a relevant qualification, such as a degree or diploma. For more details, refer to the Appendix of the 2024 AI Index Report \cite{maslej_ai_2024}{}.\\
\hline
AI Study Programs in English Penetration & The proportion of AI-related study programs offered in English in a given country. A study program, or degree program, comprises a series of courses designed to enable students to earn a relevant qualification, such as a degree or diploma. For more details, refer to the Appendix of the 2024 AI Index Report \cite{maslej_ai_2024}{}.\\
\hline
AI Talent Concentration Gender Equality Index & This indicator measures the distribution of AI talent between females and males among LinkedIn members in a given country, indicating gender equality within the AI talent pool. For more details, refer to the Appendix, Section \ref{sec:appendix-metrics}.\\
\hline
National AI Strategy Presence & This is a binary indicator showing whether a country has an AI strategy. This is a policy plan created by governments to guide the development and deployment of AI within their country. The AI Index conducted a web search to identify national AI strategies. For more details, refer to the Appendix of the 2024 AI Index Report \cite{maslej_ai_2024}{}.\\
\hline
AI Legislation Passed & The number of AI-related bills that have been enacted in a given country. For more details, refer to the Appendix of the 2024 AI Index Report \cite{maslej_ai_2024}{}.\\
\hline
AI Mentions in Legislative Proceedings & The number of times AI is mentioned in governmental and parliamentary proceedings in a given country. For more details, refer to the Appendix of the 2024 AI Index Report \cite{maslej_ai_2024}{}.\\
\hline
Social Media Share of Voice on AI & The proportion of social media discussions that mention or discuss AI attributed to a specific country compared to other countries.\\
\hline
AI Social Media Posts & The number of social media posts that mention or discuss AI in a given country.\\
\hline
AI-Related Social Media Conversations Net Sentiment & The overall sentiment of social media conversations about AI in a given country, calculated as the difference between positive and negative mentions. A net sentiment score of +100 means that all conversations are positive; a score of -100 means that all conversations are negative. \\
\hline
Parts Semiconductor Devices Exports & Total amount of exports in USD of electrical apparatus parts, including diodes, transistors, similar semiconductor devices, and photosensitive semiconductor devices for a given country. \\
\hline
Supercomputers & The number of supercomputers in a given country. \\
\hline
Compute Capacity (Rmax) & The compute capacity, measured as Rmax (GFlops), representing the performance of supercomputers in a given country. \\
\hline
Internet Speed & This indicator represents the median internet download speed in Mbps in a given country. \\

\end{longtable}
\label{tab:indicator_list_description}
\end{center}

\pagebreak

\section{INDICATORS INCLUDED IN SUB-INDICES}
\label{sec:appendix-subindices-indicators}

\begin{table}[h!]
\small
\centering
\caption{Innovation Index Indicators}
\label{tab:innovation_index_indicators}
\begin{tabular}{|l|}
\hline
\textbf{Key indicators included} \\ \hline
AI Journal Publications          \\ \hline
AI Journal Citations             \\ \hline
AI Conference Publications       \\ \hline
AI Conference Citations          \\ \hline
AI Patent Grants                 \\ \hline
Notable Machine Learning Models  \\ \hline
Foundation Models                \\ \hline
Foundation Models Datasets       \\ \hline
Foundation Models Applications   \\ \hline
AI GitHub Projects               \\ \hline
AI GitHub Projects Stars         \\ \hline
Parts Semiconductor Devices Exports \\ \hline
Supercomputers                   \\ \hline
Compute Capacity (Rmax)          \\ \hline
Internet Speed                   \\ \hline
\end{tabular}
\end{table}

\begin{table}[h!]
\small
\centering
\caption{Economic Competitiveness Index Indicators}
\label{tab:economic_competitiveness_index_indicators}
\begin{tabular}{|l|}
\hline
\textbf{Key indicators included}          \\ \hline
Total AI Private Investment               \\ \hline
Total AI Merger/Acquisition Investment    \\ \hline
Total AI Minority Stake Investment        \\ \hline
Total AI Public Offering Investment       \\ \hline
Newly Funded AI Companies                 \\ \hline
AI Hiring Rate YoY Ratio                  \\ \hline
Relative AI Skill Penetration             \\ \hline
AI Talent Concentration                   \\ \hline
Net Migration Flow of AI Skills           \\ \hline
\end{tabular}
\end{table}

\begin{table}[h!]
\small
\centering
\caption{Policy, Governance, and Public Engagement Index Indicators}
\label{tab:policy_governance_index_indicators}
\begin{tabular}{|l|}
\hline
\textbf{Key indicators included}             \\ \hline
National AI Strategy Presence                \\ \hline
AI Legislation Passed                        \\ \hline
AI Mentions in Legislative Proceedings       \\ \hline
Social Media Share of Voice on AI            \\ \hline
AI Social Media Posts                        \\ \hline
AI-Related Social Media Conversations Net Sentiment \\ \hline
\end{tabular}
\end{table}

\pagebreak

\section{Weighting: Budget Allocation Method}
\label{sec:appendix-weighting}

In assigning weights to pillars and indicators, careful consideration was given to data availability, quality, and the scope of coverage across countries to ensure a reliable representation of AI vibrancy. Indicators with limited coverage—such as \textit{Academia-Industry Model Production Concentration}, \textit{Open Access Foundation Models}, and \textit{AI Job Postings (\% of Total)}—were assigned a weight of zero, as their availability (only covering 33\%, 36\%, and 39\% of countries respectively in 2023) undermines their utility in cross-country comparisons. Data enhancement efforts are underway for these indicators to improve future analyses. Similarly, weights for indicators like \textit{AI Hiring Rate YoY Ratio}, \textit{Relative AI Skill Penetration}, \textit{AI Talent Concentration}, and \textit{Net Migration Flow of AI Skills} were reduced due to partial data coverage (69\% in 2023). 
Further, foundational AI indicators, including \textit{Foundation Models} (count), \textit{Foundation Model Applications}, and \textit{Foundation Model Datasets}, were assigned a moderate weight of 3 to balance their importance with the underrepresentation of some countries due to the predominance of English-language data sources, with ongoing efforts to enhance data accuracy across more countries. Pillar weights were also adjusted to reflect data quality and relevance. For example, the \textit{Responsible AI} pillar was assigned a low weight of 2 due to potential overlap with the \textit{R\&D} pillar, though the search for additional country-specific metrics continues. The \textit{Economy} pillar’s weight was adjusted to 8 to indicate a need for more indicators on AI technology adoption, while the Education pillar, weighted at 2, requires further development to mitigate English-language biases.
Other adjustments include reducing the weight of the \textit{Diversity} pillar to 1, due to its limited indicators and low data coverage, and the \textit{Policy and Governance} pillar to 4, addressing the gap in both qualitative and quantitative metrics, such as AI-dedicated government budgets. The \textit{Public Opinion} pillar received a weight of 2, acknowledging its lesser relevance compared to pillars like \textit{Policy and Governance}, while the \textit{Infrastructure} pillar was weighted at 6. Although critical, this adjustment reflects missing metrics on infrastructure specifics, like data center availability, with plans to expand data collection underway. This calibration provides a comprehensive view of global AI vibrancy, accounting for some data limitations. The approach prevents skewed assessments from data gaps or overemphasis, resulting in a more balanced analysis.

\begin{table}[h!]
\small
\centering
\caption{Pillar Weights}
\label{tab:pillar_weights}
\begin{tabular}{|l|c|c|}
\hline
\textbf{Pillar} & \textbf{Pillar Weight (0 - 10)} & \textbf{Pillar Weight (\%)} \\ \hline
Research and Development & 10 & 28.57\% \\ \hline
Responsible AI & 2 & 5.71\% \\ \hline
Economy & 8 & 22.86\% \\ \hline
Education & 2 & 5.71\% \\ \hline
Diversity & 1 & 2.86\% \\ \hline
Policy and Governance & 4 & 11.43\% \\ \hline
Public Opinion & 2 & 5.71\% \\ \hline
Infrastructure & 6 & 17.14\% \\ \hline
\end{tabular}
\end{table}

\begin{table}[h!]
\small
\centering
\caption{Indicator Weights}
\label{tab:indicator_weights}
\begin{tabular}{|l|c|c|}
\hline
\textbf{Indicator} & \textbf{Indicator Weight (0 - 10)} & \textbf{Indicator Weight (\%)} \\ \hline
AI Journal Publications & 8 & 11.43\% \\ \hline
AI Journal Citations & 8 & 11.43\% \\ \hline
AI Conference Publications & 6 & 8.57\% \\ \hline
AI Conference Citations & 7 & 10.00\% \\ \hline
AI Patent Grants & 8 & 11.43\% \\ \hline
Notable Machine Learning Models & 9 & 12.86\% \\ \hline
Academia-Industry Model Production Concentration & 0 & 0.00\% \\ \hline
Foundation Models & 3 & 4.29\% \\ \hline
Foundation Models Datasets & 3 & 4.29\% \\ \hline
Foundation Models Applications & 3 & 4.29\% \\ \hline
Open Access Foundation Models & 0 & 0.00\% \\ \hline
AI GitHub Projects & 7 & 10.00\% \\ \hline
AI GitHub Projects Stars & 8 & 11.43\% \\ \hline
FAccT Conference Submissions on RAI Topics & 7 & 15.22\% \\ \hline
NeurIPS Conference Submissions on RAI Topics & 10 & 21.74\% \\ \hline
ICML Conference Submissions on RAI Topics & 8 & 17.39\% \\ \hline
ICLR Conference Submissions on RAI Topics & 7 & 15.22\% \\ \hline
AIES Conference Submissions on RAI Topics & 6 & 13.04\% \\ \hline
AAAI Conference Submissions on RAI Topics & 8 & 17.39\% \\ \hline
Total AI Private Investment & 10 & 15.87\% \\ \hline
Total AI Merger/Acquisition Investment & 9 & 14.29\% \\ \hline
Total AI Minority Stake Investment & 7 & 11.11\% \\ \hline
Total AI Public Offering Investment & 7 & 11.11\% \\ \hline
Newly Funded AI Companies & 9 & 14.29\% \\ \hline
AI Hiring Rate YoY Ratio & 6 & 9.52\% \\ \hline
Relative AI Skill Penetration & 3 & 4.76\% \\ \hline
AI Talent Concentration & 6 & 9.52\% \\ \hline
AI Job Postings (\% of Total) & 0 & 0.00\% \\ \hline
Net Migration Flow of AI Skills & 6 & 9.52\% \\ \hline
AI Study Programs in English & 6 & 46.15\% \\ \hline
AI Study Programs in English Penetration & 7 & 53.85\% \\ \hline
AI Talent Concentration Gender Equality Index & 10 & 100.00\% \\ \hline
National AI Strategy Presence & 10 & 38.46\% \\ \hline
AI Legislation Passed & 10 & 38.46\% \\ \hline
AI Mentions in Legislative Proceedings & 6 & 23.08\% \\ \hline
Social Media Share of Voice on AI & 8 & 34.78\% \\ \hline
AI Social Media Posts & 6 & 26.09\% \\ \hline
AI-Related Social Media Conversations Net Sentiment & 9 & 39.13\% \\ \hline
Parts Semiconductor Devices Exports & 10 & 27.03\% \\ \hline
Supercomputers & 9 & 24.32\% \\ \hline
Compute Capacity (Rmax) & 10 & 27.03\% \\ \hline
Internet Speed & 8 & 21.62\% \\ \hline
\end{tabular}
\end{table}

\pagebreak

\section{Metrics Calculation}
\label{sec:appendix-metrics}

\subsection{Academia-Industry Model Production Concentration}
\label{sec:appendix-model-production}
The \textit{Academia-Industry Model Production Concentration} indicator quantifies the balance between the number of notable machine learning models produced by academia and industry within each country and year. The indicator is calculated using the inverted Herfindahl-Hirschman Index (HHI), a measure commonly used to assess market concentration.

\begin{equation}
    \text{Inverted HHI} = 1 - (\alpha^2 + \iota^2)
\end{equation}

where:
\begin{itemize}
    \item $\alpha$ is the proportion of models produced only by academia.
    \item $\iota$ is the proportion of models produced only by industry.
\end{itemize}

A higher normalized value indicates a more balanced model production environment, suggesting a favorable condition for AI development due to collaboration and diversity between academia and industry. Conversely, a lower normalized value indicates a concentration in model production by either academia or industry, suggesting less diversity.

\subsection{AI Talent Concentration Gender Equality Index}
\label{sec:appendix-ai-talent-concentration-gender}

The \textit{AI Talent Concentration Gender Equality Index} measures how evenly the proportion of LinkedIn members who are AI talent are distributed between female and males in a country $k$.\footnote{Refer to the 2024 AI Index Report \cite{maslej_ai_2024} for detailed information on the definition and calculation of the metric.} The index is calculated using the following formula:

\begin{equation}
    \frac{\min(\text{Female}_{k}, \text{Male}_{k})}{\max(\text{Female}_{k}, \text{Male}_{k})}
\end{equation}

where:
\begin{itemize}
    \item Female$_k$ is the AI talent concentration of females.
    \item Male$_k$ is the AI talent concentration of males.
\end{itemize}

The index ranges from 0 to 1, where 1 means perfect equality and 0 means complete inequality. A higher value of the index means a more balanced distribution of AI talent between females and males, reflecting greater gender equality in AI talent concentration within that country.

\newpage
\section{Data Coverage}
\label{sec:appendix-coverage}

\begin{table}[h!]
\caption{Data Coverage by Country and Year}
\centering
\begin{tabular}{lrrrrrrr}
\toprule
\textbf{Country} & \textbf{2017} & \textbf{2018} & \textbf{2019} & \textbf{2020} & \textbf{2021} & \textbf{2022} & \textbf{2023} \\
\midrule
Australia &  60\% &  62\% &  81\% &  88\% &  88\% &  88\% &  86\% \\
Austria &  48\% &  45\% &  64\% &  76\% &  74\% &  74\% &  71\% \\
Belgium &  50\% &  57\% &  71\% &  81\% &  81\% &  86\% &  88\% \\
Brazil &  50\% &  57\% &  76\% &  86\% &  86\% &  86\% &  83\% \\
Canada &  60\% &  62\% &  79\% &  90\% &  98\% & 100\% &  98\% \\
China &  52\% &  52\% &  67\% &  76\% &  86\% &  86\% &  83\% \\
Denmark &  50\% &  52\% &  69\% &  79\% &  79\% &  79\% &  86\% \\
Estonia &  45\% &  45\% &  60\% &  69\% &  69\% &  69\% &  67\% \\
Finland &  55\% &  57\% &  74\% &  83\% &  83\% &  83\% &  93\% \\
France &  57\% &  64\% &  88\% &  88\% &  88\% &  90\% &  98\% \\
Germany &  57\% &  62\% &  81\% &  90\% &  98\% &  98\% &  98\% \\
India &  55\% &  62\% &  76\% &  86\% &  86\% &  86\% &  86\% \\
Ireland &  55\% &  62\% &  74\% &  86\% &  83\% &  83\% &  83\% \\
Israel &  55\% &  55\% &  74\% &  88\% &  90\% &  90\% &  90\% \\
Italy &  55\% &  62\% &  79\% &  88\% &  88\% &  88\% &  86\% \\
Japan &  52\% &  50\% &  64\% &  76\% &  74\% &  74\% &  71\% \\
Luxembourg &  50\% &  52\% &  69\% &  79\% &  71\% &  83\% &  81\% \\
Malaysia &  45\% &  45\% &  60\% &  69\% &  69\% &  69\% &  67\% \\
Mexico &  50\% &  45\% &  60\% &  57\% &  69\% &  69\% &  67\% \\
Netherlands &  52\% &  60\% &  79\% &  86\% &  83\% &  86\% &  86\% \\
New Zealand &  48\% &  60\% &  71\% &  81\% &  81\% &  81\% &  79\% \\
Norway &  55\% &  57\% &  74\% &  86\% &  83\% &  86\% &  83\% \\
Poland &  52\% &  50\% &  64\% &  74\% &  74\% &  74\% &  71\% \\
Portugal &  50\% &  52\% &  69\% &  79\% &  79\% &  81\% &  79\% \\
Russia &  45\% &  48\% &  60\% &  69\% &  69\% &  79\% &  55\% \\
Saudi Arabia &  36\% &  36\% &  62\% &  71\% &  71\% &  74\% &  57\% \\
Singapore &  57\% &  62\% &  76\% &  83\% &  83\% &  86\% &  93\% \\
South Africa &  55\% &  57\% &  69\% &  81\% &  79\% &  81\% &  79\% \\
South Korea &  50\% &  48\% &  64\% &  76\% &  86\% &  83\% &  81\% \\
Spain &  55\% &  62\% &  79\% &  88\% &  88\% &  88\% &  95\% \\
Sweden &  55\% &  60\% &  79\% &  88\% &  86\% &  88\% &  95\% \\
Switzerland &  55\% &  62\% &  81\% &  88\% &  90\% &  90\% &  98\% \\
Turkey &  26\% &  38\% &  40\% &  62\% &  62\% &  62\% &  60\% \\
United Arab Emirates &  45\% &  48\% &  69\% &  79\% &  79\% &  83\% &  88\% \\
United Kingdom &  62\% &  64\% &  81\% &  90\% & 100\% & 100\% &  98\% \\
United States &  62\% &  64\% &  90\% & 100\% & 100\% & 100\% &  98\% \\
\bottomrule
\end{tabular}
\label{tab:data_coverage_countries}
\end{table}

\begin{table}[h!]
\caption{Data Coverage by Indicator and Year}
\centering
\small
\begin{tabular}{lrrrrrrr}
\toprule
\textbf{Indicator} & \textbf{2017} & \textbf{2018} & \textbf{2019} & \textbf{2020} & \textbf{2021} & \textbf{2022} & \textbf{2023} \\
\midrule
Academia-Industry Model Production Concentration & 31\% & 28\% & 25\% & 25\% & 17\% & 28\% & 33\% \\
AI Job Postings (\% of Total) & 14\% & 39\% & 39\% & 39\% & 39\% & 39\% & 39\% \\
AI Mentions in Legislative Proceedings & 89\% & 89\% & 89\% & 89\% & 89\% & 89\% & 89\% \\
AI-Related Social Media Conversations Net Sentiment & N/A & N/A & N/A & 100\% & 100\% & 100\% & 100\% \\
AI Social Media Posts & N/A & N/A & N/A & 100\% & 100\% & 100\% & 100\% \\
AI Study Programs in English Penetration & 92\% & 92\% & 92\% & 92\% & 92\% & 92\% & 92\% \\
AI Talent Concentration & 69\% & 69\% & 69\% & 69\% & 69\% & 69\% & 69\% \\
AI Talent Concentration Gender Equality Index & 69\% & 69\% & 69\% & 69\% & 69\% & 69\% & 69\% \\
Compute Capacity (Rmax) & 72\% & 69\% & 69\% & 69\% & 72\% & 75\% & 75\% \\
Internet Speed & N/A & N/A & N/A & 97\% & 97\% & 97\% & 97\% \\
Total AI Merger/Acquisition Investment & 92\% & 97\% & 97\% & 97\% & 97\% & 100\% & 94\% \\
Total AI Minority Stake Investment & 92\% & 97\% & 97\% & 97\% & 97\% & 100\% & 94\% \\
National AI Strategy Presence & 100\% & 100\% & 100\% & 100\% & 100\% & 100\% & 100\% \\
Net Migration Flow of AI Skills & N/A & N/A & 69\% & 69\% & 69\% & 69\% & 69\% \\
AI Legislation Passed & 100\% & 100\% & 100\% & 100\% & 100\% & 100\% & 100\% \\
AI Study Programs in English & 94\% & 94\% & 94\% & 94\% & 94\% & 94\% & 94\% \\
Foundation Models & N/A & N/A & 6\% & 6\% & 19\% & 22\% & 42\% \\
Foundation Models Applications & N/A & N/A & 6\% & 6\% & 19\% & 22\% & 42\% \\
Foundation Models Datasets & N/A & N/A & 6\% & 6\% & 19\% & 22\% & 42\% \\
AI GitHub Projects & 100\% & 100\% & 100\% & 100\% & 100\% & 100\% & 100\% \\
AI GitHub Projects Stars & 100\% & 100\% & 100\% & 100\% & 100\% & 100\% & 100\% \\
Newly Funded AI Companies & 94\% & 97\% & 97\% & 97\% & 97\% & 100\% & 94\% \\
Notable Machine Learning Models & 100\% & 100\% & 100\% & 100\% & 100\% & 100\% & 100\% \\
AAAI Conference Submissions on RAI Topics & N/A & N/A & 100\% & 100\% & 100\% & 100\% & 100\% \\
AIES Conference Submissions on RAI Topics & N/A & N/A & 100\% & 100\% & 100\% & 100\% & 100\% \\
FAccT Conference Submissions on RAI Topics & N/A & N/A & 100\% & 100\% & 100\% & 100\% & 100\% \\
ICLR Conference Submissions on RAI Topics & N/A & N/A & 100\% & 100\% & 100\% & 100\% & 100\% \\
ICML Conference Submissions on RAI Topics & N/A & N/A & 100\% & 100\% & 100\% & 100\% & 100\% \\
NeurIPS Conference Submissions on RAI Topics & N/A & N/A & 100\% & 100\% & 100\% & 100\% & 100\% \\
Supercomputers & 72\% & 69\% & 69\% & 69\% & 72\% & 75\% & 75\% \\
AI Conference Citations & 100\% & 100\% & 100\% & 100\% & 100\% & 100\% & 100\% \\
AI Conference Publications & 100\% & 100\% & 100\% & 100\% & 100\% & 100\% & 100\% \\
AI Journal Citations & 100\% & 100\% & 100\% & 100\% & 100\% & 100\% & 100\% \\
AI Journal Publications & 100\% & 100\% & 100\% & 100\% & 100\% & 100\% & 100\% \\
AI Patent Grants & 97\% & 97\% & 97\% & 97\% & 97\% & 97\% & 97\% \\
Parts Semiconductor Devices Exports & 97\% & 97\% & 97\% & 97\% & 97\% & 97\% & N/A \\
Open Access Foundation Models & N/A & N/A & 3\% & 3\% & 17\% & 19\% & 36\% \\
Total AI Private Investment & 92\% & 97\% & 97\% & 97\% & 97\% & 100\% & 94\% \\
Total AI Public Offering Investment & 92\% & 97\% & 97\% & 97\% & 97\% & 100\% & 94\% \\
AI Hiring Rate YoY Ratio & N/A & 69\% & 69\% & 69\% & 69\% & 69\% & 69\% \\
Relative AI Skill Penetration & 11\% & 36\% & 44\% & 50\% & 39\% & 56\% & 67\% \\
Social Media Share of Voice on AI & N/A & N/A & N/A & 100\% & 100\% & 100\% & 100\% \\
\bottomrule
\end{tabular}

\label{tab:data_coverage_indicators}
\end{table}

\newpage
\section{Results of Country Rankings}
\label{sec:appendix-results}

\subsection{Global AI Vibrancy}
\label{sec:appendix-global_rankings_all_countries}

\begin{figure}[H]
    \centering
    \includegraphics[width=1\linewidth]{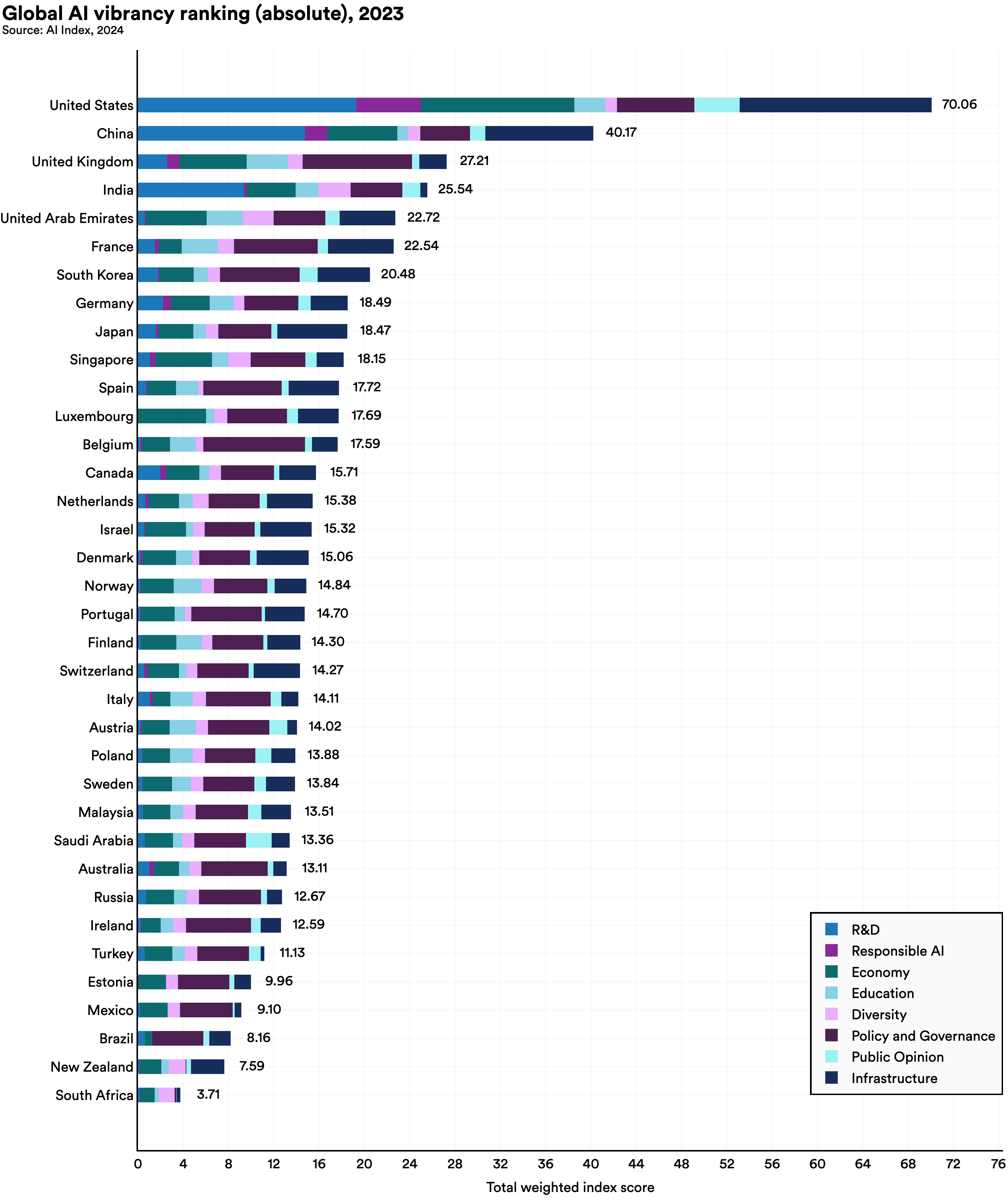}
    \caption{}
    \label{fig: global_ai_vibrancy_ranking_all}
\end{figure}

\begin{figure}[H]
    \centering
    \includegraphics[width=1\linewidth]{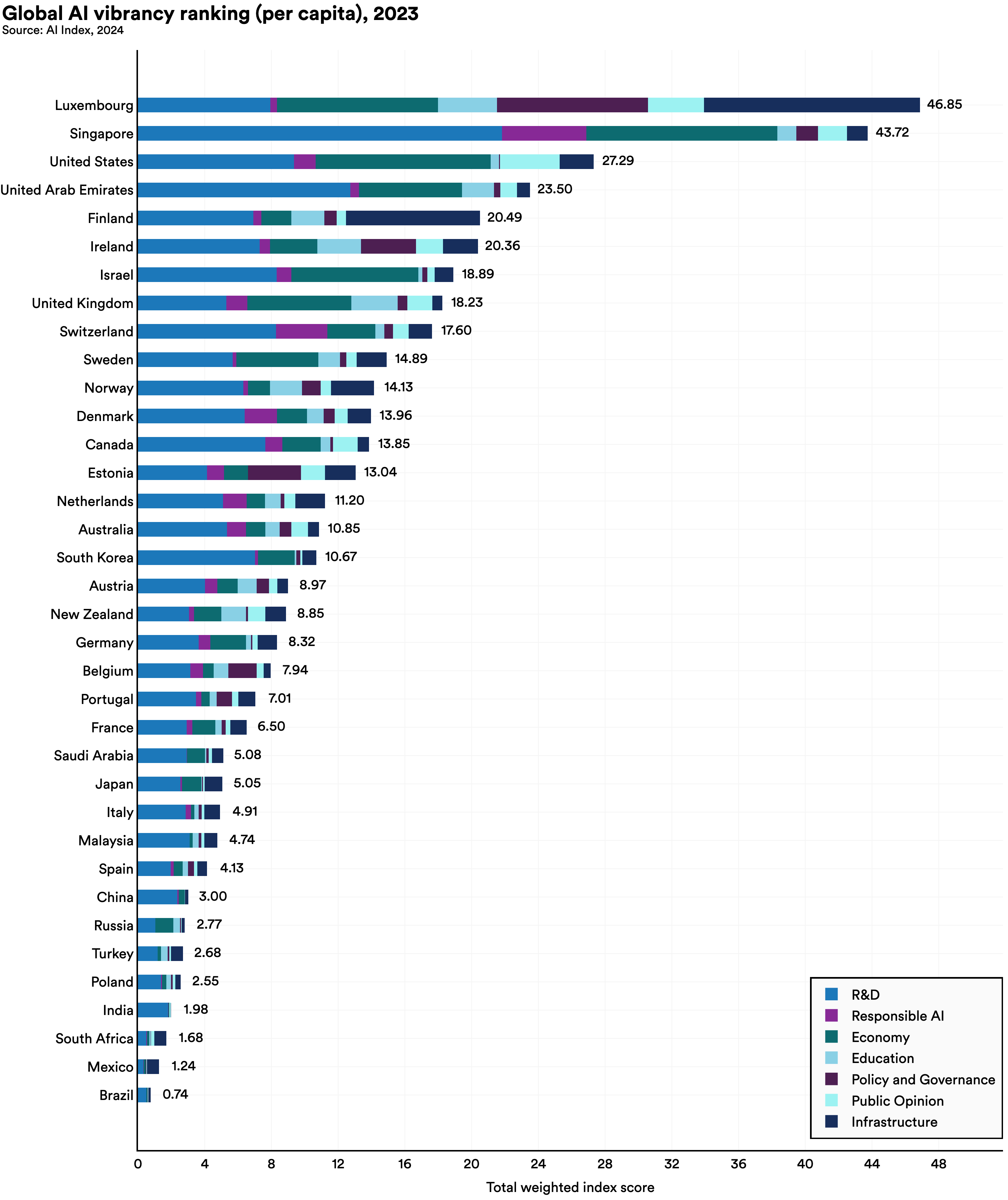}
    \caption{}
    \label{fig: global_ai_vibrancy_ranking_all_percapita}
\end{figure}

\subsection{Sub-Indices}
\label{sec:appendix-sub_indices_rankings_all_countries}

\begin{figure}[H]
    \centering
    \includegraphics[width=1\linewidth]{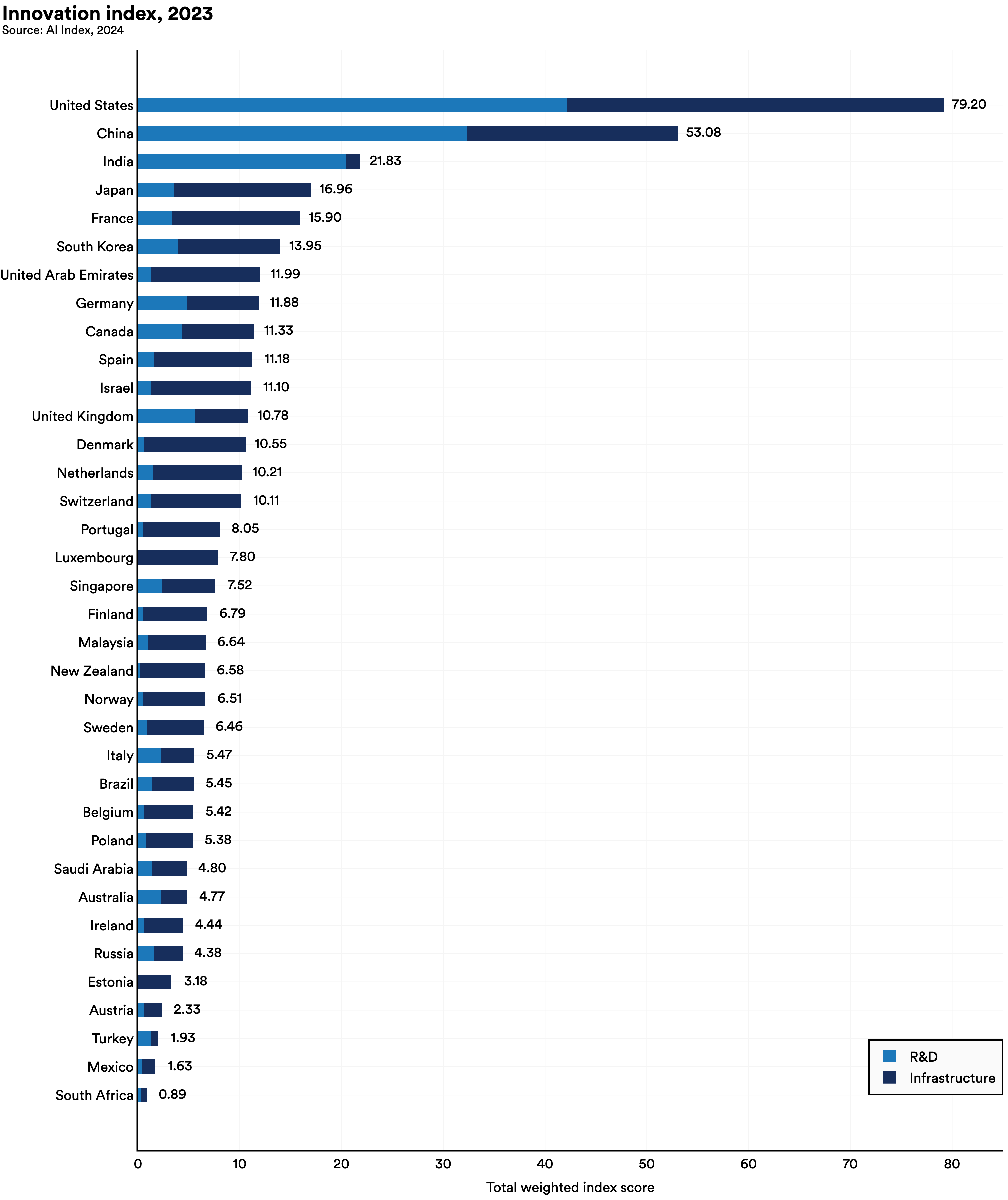}
    \caption{}
    \label{fig: innovation_ranking_all_countries_2023}
\end{figure}

\begin{figure}[H]
    \centering
    \includegraphics[width=1\linewidth]{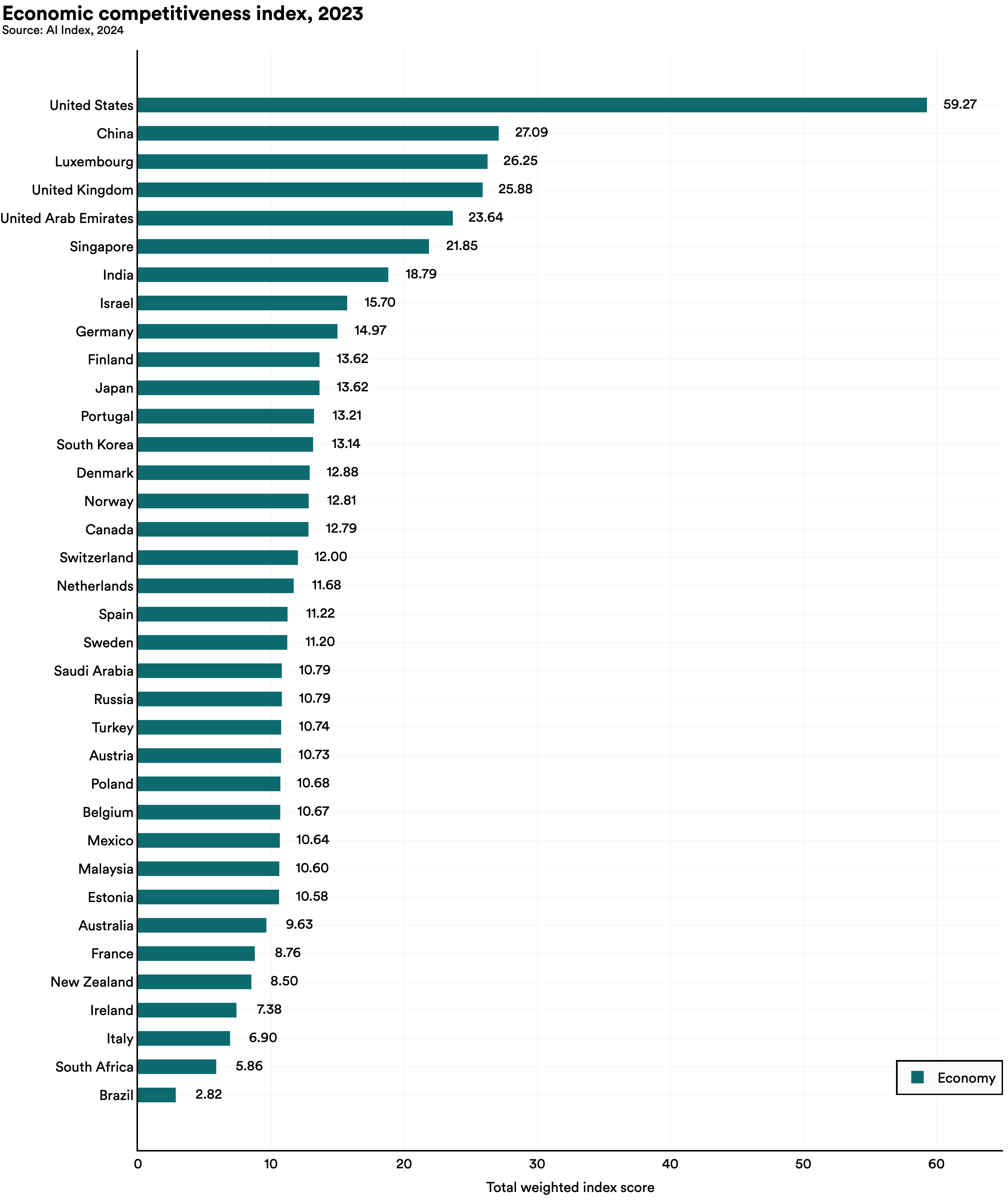}
    \caption{}
    \label{fig: economy_ranking_all_countries_2023}
\end{figure}

\begin{figure}[H]
    \centering
    \includegraphics[width=1\linewidth]{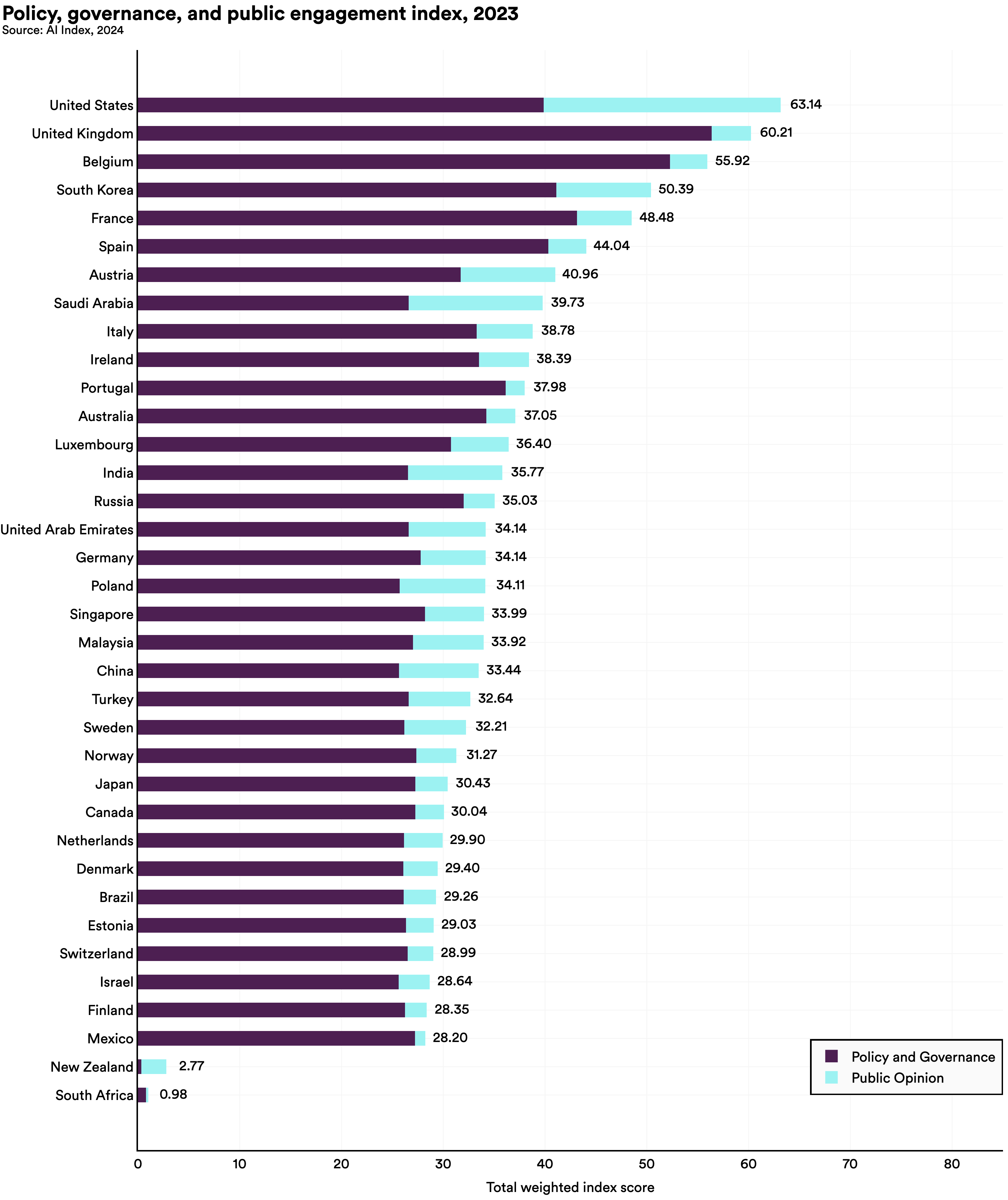}
    \caption{}
    \label{fig: policy_governance_public_opinion_ranking_all_countries_2023}
\end{figure}

\end{appendices}

\end{document}